\documentclass[12pt,prd,nofootinbib,superscriptaddress,amsmath,amssymb,aps,reprint]{revtex4-1}
\usepackage{xparse, physics, graphicx, dcolumn, bm, amsmath, amssymb, mathtools, cancel, dsfont, enumitem, braket, tikz, color}
\usepackage{soul,xcolor}
\usepackage[linktocpage=true]{hyperref}
\usepackage[all]{hypcap}
\usepackage{cleveref}
\usepackage{siunitx}
\usepackage{cleveref}
\usepackage{comment}
\usepackage{float}
\usepackage{subcaption}
\usepackage{afterpage}
\usepackage[normalem]{ulem}
\usepackage{booktabs,tabularx} 
\usepackage{caption}           

\newcommand{\stkout}[1]{\ifmmode\text{\sout{\ensuremath{#1}}}\else\sout{#1}\fi}

\crefname{equation}{Eq}{Eq's.} 
\crefrangelabelformat{equation}{(#3#1#4--#5#2#6)}

\newcommand{\MPl}{M_\text{Pl}}
\newcommand{\defeq}{\triangleq}
\newcommand{\GammaNGBWall}{{\Gamma_\text{wall}^\text{NGB}}}

\setstcolor{violet}

\newcommand{\cmt}[1]{}

\begin{document}
\title{Searching Stochastic Gravitational Wave Background Landscape Across Frequency Bands}

\author{Yunjia Bao}
\email{yunjia.bao@uchicago.edu}
\affiliation{Department of Physics, University of Chicago, Chicago, IL, 60637, USA}
\affiliation{Enrico Fermi Institute, University of Chicago, Chicago, IL, 60637, USA}
\affiliation{Kavli Institute for Cosmological Physics, University of Chicago, Chicago, IL 60637, USA}
\affiliation{Leinweber Institute for Theoretical Physics, University of Chicago, Chicago, IL 60637, USA}
\author{Töre Boybeyi}
\email{boybe001@umn.edu}
\affiliation{School of Physics and Astronomy, University of Minnesota, Minneapolis, MN 55455, USA}
\author{Vuk Mandic}
\email{vuk@umn.edu}
\affiliation{School of Physics and Astronomy, University of Minnesota, Minneapolis, MN 55455, USA}
\author{Lian-Tao Wang}
\email{liantaow@uchicago.edu}
\affiliation{Department of Physics, University of Chicago, Chicago, IL, 60637, USA}
\affiliation{Enrico Fermi Institute, University of Chicago, Chicago, IL, 60637, USA}
\affiliation{Kavli Institute for Cosmological Physics, University of Chicago, Chicago, IL 60637, USA}
\affiliation{Leinweber Institute for Theoretical Physics, University of Chicago, Chicago, IL 60637, USA}
\affiliation{HEP Division, Argonne National Laboratory, 9700 Cass Ave., Argonne, IL 60439, USA}
\begin{abstract}
    Gravitational wave (GW) astrophysics is entering a multi-band era with upcoming GW detectors, enabling detailed mapping of the stochastic GW background across vast frequencies. We highlight this potential via a new physics scenario: hybrid topological defects from a two-step phase transition separated by inflation. We develop a general pipeline to analyze experimental exclusions and apply it to this model. The model offers a possible explanation of the pulsar timing array signal at low frequencies, and future experiments (LISA/Cosmic Explorer/Einstein Telescope) will confirm or rule it out via the higher-frequency probes, showcasing the power of multi-band constraints.
\end{abstract}

\maketitle

\section{Introduction}
Gravitational wave (GW) astrophysics is an emerging window into the early universe. Current detectors primarily operate in two bands: (1) ground-based interferometric observatories, such as the LIGO-Virgo-KAGRA (LVK) collaboration \cite{LIGOScientific:2014pky, VIRGO:2014yos, KAGRA:2018plz}, operating in $\sim \SIrange{E1}{E3}{\Hz}$, and (2) pulsar timing arrays (PTAs), such as NANOGrav \cite{NANOGrav:2023gor}, EPTA and InPTA \cite{EPTA:2023fyk}, CPTA \cite{xu2023searching}, IPTA \cite{Antoniadis:2022pcn}, in $\sim \SIrange{E-9}{E-7}{\Hz}$. The LVK network has provided upper bounds on the isotropic stochastic GW background (SGWB) since its O1 run \cite{abbott2017upper} and continues to do so in its O3 run \cite{abbott2021upper} and O4a run \cite{LIGOScientific:2025bgj}. With the O4 run underway and the O5 run scheduled, we expect further insight into the early-universe dynamics from ground-based detectors \cite{LIGOScientific:2025kry}. The recent consistent detections of a GW background in the nHz band by multiple collaborations \cite{NANOGrav:2023gor, antoniadis2023second, reardon2023search, xu2023searching} spurred great excitement, demonstrating the maturity of low-frequency SGWB mapping in the 2020s. Especially striking is the surprisingly large amplitude of the signal, motivating investigations from both the theoretical and observational frontiers. 

In the upcoming decade, many GW detectors shall join the current flagships, broadly mapping the GW spectrum across frequencies. The 3\textsuperscript{rd}-generation ground-based detectors, such as Cosmic Explorer (CE) \cite{LIGOScientific:2016wof, Reitze:2019iox} and Einstein Telescope (ET) \cite{Punturo:2010zz, ET:2019dnz}, aim to uncover signals up to $10^{-3}$ to $10^{-4}$ fainter than current sensitivity for the SGWB. In the nHz window, the Square Kilometer Array can probe $10^{-5}$ to $10^{-6}$ below current sensitivity in energy density \cite{Janssen:2014dka, Weltman:2018zrl}. Beyond these two bands, a plethora of space-borne detectors, such as LISA \cite{Baker:2019nia, Caldwell:2019vru}, DECIGO \cite{Kawamura:2020pcg, Isoyama:2018rjb}, BBO \cite{Corbin:2005ny, Harry:2006fi}, TianQin \cite{TianQin:2015yph, TianQin:2020hid}, Taiji \cite{Hu:2017mde, Luo:2021qji}, shall operate in intermediate frequencies, bridging the gap between PTA and LVK frequency bands. These future experiments prompt a unified study of the SGWB across frequencies, as such opportunities rapidly become feasible. In fact, some multi-band search strategies for new physics, such as primordial black hole production from large curvature perturbation, have been investigated \cite{Wang:2019kaf, Zhao:2022kvz, Inomata:2023zup, Cang:2023ysz, Clesse:2024epo, Gouttenoire:2025jxe}.

A cosmogenic SGWB is particularly appealing. If detected, this will serve as a trailblazing probe of the early universe, on which we have limited information. One cosmogenic source is topological defects formed in the early universe, such as cosmic strings~\cite{Vilenkin:1981iu, Vilenkin:1981zs, Hogan:1984is, Sakellariadou:1990ne, Damour:2000wa, Siemens:2006yp, Lorenz:2010sm, Olmez:2010bi, Sousa:2013aaa, LIGOScientific:2013tfe, Blanco-Pillado:2017oxo, Ringeval:2017eww, Cui:2017ufi, LIGOScientific:2017ikf, Cui:2018rwi, Cui:2019kkd, Auclair:2019wcv, Blasi:2020mfx, Ellis:2020ena, Sousa:2020sxs, Figueroa:2020lvo, Co:2021lkc, Gorghetto:2021fsn, Buchmuller:2021mbb, Chang:2021afa, Boileau:2021gbr, Gouttenoire:2021wzu, Gouttenoire:2021jhk, Hindmarsh:2022awe, Ferrer:2023uwz, Auclair:2023brk, Baeza-Ballesteros:2023say, Kume:2024adn, Fedderke:2025sic}, metastable domain walls~\cite{Vachaspati:1984gt, Gleiser:1998na, Hiramatsu:2010yz, Hiramatsu:2013qaa, Kawasaki:2011vv, Kamada:2015iga, Nakayama:2016gxi, Ferreira:2022zzo, Bai:2023cqj, Ge:2023rce, Kitajima:2023cek, An:2023idh}, and other hybrid defect networks~\cite{Martin:1996cp, Babichev:2004gy, Dunsky:2021tih, Lazarides:2022jgr, Roshan:2024qnv, Chitose:2023dam, Maji:2025yms}. 

In this study, we considered a particular scenario in which a two-step phase transition — first producing cosmic strings, then domain walls — is separated by an inflationary epoch \cite{Bao:2024bws}. This new physics scenario is well-motivated theoretically with interesting GW signatures. We use this model as a testbed to explore how current and near-future GW observatories constrain its parameters, emphasizing the complementarities and significance of multi-band GW observations. We also use this model to offer a possible explanation of the NANOGrav signal and show how future experiments could corroborate or refute this interpretation.

The paper is organized as follows: \Cref{sec:theoreticalModels} discusses the new physics scenario and its GW signal. Depending on whether the boundary cosmic string is gauged (\cref{sec:gaugeModel}) or global (\cref{sec:globalModel}), the GW spectrum will slightly differ. In \cref{sec:methodology}, we detail how the Bayesian analysis is done and how current exclusion and projected sensitivity reaches are modeled with more details offered in \cref{S_App1,S_App2,app:S_SMBHB} of Supplemental Materials.
\footnote{Supplemental materials contain details of the sensitivity modeling and additional discussion of the string network modeling, which includes Refs.~\cite{maggiore2008gravitational, Romano:2016dpx, Prince:2002hp, cornish2001space, punturo2010third, amaro2017laser, Robson:2018ifk, Criswell:2024hfn, Thrane:2013oya, moore2014gravitational, schmitz2020new, FitzAxen:2018vdt, Stott:2016loe, burke2019astrophysics, Sesana:2013wja, NANOGrav:2023hfp, afzal2023nanograv, Phinney:2001di}.}
Results from the Bayesian analysis are presented in \cref{sec:results}, highlighting \cref{fig:1a,fig:1b,fig:2a,fig:2b}. We conclude in \cref{sec:conclusion}. 
\section{Theoretical Models \label{sec:theoreticalModels}}
In this study, we consider the SGWB from hybrid topological defects, whose signal covers a broad frequency range following Ref.~\cite{Bao:2024bws}. Specifically, we assume that two phase transitions occur in the early universe: one produces cosmic strings, and the other produces domain walls. Topology dictates that cosmic strings bound domain walls. If the typical radius of boundary strings is small, the dominant GW signal comes from the string network, and the main distinction between this string-wall network and a pure string network is the IR cutoff frequency \cite{Dunsky:2021tih}. However, if an epoch of inflation occurs between the two phase transitions, the boundary cosmic strings are pushed outside the comoving horizon, enlarging the typical wall size. This enhances the GW signal from domain walls, producing a characteristic peak atop the otherwise flat string-sourced GW spectrum \cite{Bao:2024bws}. The GW spectrum obtained in Ref.~\cite{Bao:2024bws} is based largely on scaling arguments. Although detailed simulations are needed for precise predictions, the results should remain qualitatively unchanged. For a detailed timeline for the defect network's evolution, see Fig.~2 of Ref.~\cite{Bao:2024bws}. The peak frequency and amplitude of the wall GW spectrum are partly determined by the Hubble scale $H_\text{re}$ when most boundary strings re-enter the horizon. Since this scale depends jointly on the correlation length of the string-forming scalar field from the Kibble-Zurek mechanism \cite{kibble1976topology, zurek1985cosmological} and the number of $e$-foldings of inflation, $H_\text{re}$ can be treated as a free parameter with a wide theoretical prior. This flexibility allows the wall signals to peak at arbitrary frequencies. In what follows, we present two particular scenarios: one in which walls are bounded by gauge strings and the other by global strings. 

\subsection{New Physics Scenario 1: Walls Bounded by Gauge Cosmic Strings \label{sec:gaugeModel}}
This model includes three parameters: the domain wall tension $\sigma$ (mass dimension 3), the re-entry Hubble parameter $H_\text{re}$ (mass dimension 1), and the vacuum expectation value (VEV) $v_2$ (mass dimension 1) of the string-forming scalar field.%
\footnote{Parameterizing using $v_2$ is to compare fairly with the global-string scenario. The string tension $\mu$ (mass dimension 2) relates to $v_2$ by $\mu = \pi v_2^2$.} 
These parameters come from different physics. While $v_2$ and $\sigma$ are determined by the two phase transitions respectively, $H_\text{re}$ is determined by the inflationary dynamics. In particular, the re-entry comoving Hubble scale is exponentially sensitive to the number of $e$-foldings from phase transitions to the end of inflation, and even if $H_\text{re}$ has power-law dependence on parameters sensitive to the phase transition, this exponential dependence on allows us to treat $H_\text{re}$ as essentially an independent parameter from $v_2$ or $\sigma$. Hence, all three parameters can be treated as independent. 

The GW spectrum consists of two components: the wall spectrum and the string spectrum. The former only depends on $\sigma$ and $H_\text{re}$, assuming that the wall can predominantly decay into gravitational waves with a decay rate $\Gamma_\text{wall} \defeq \pi \sigma/\MPl^2$. Here, we report the GW fractional energy when $H = \Gamma_\text{wall}$
\begin{equation}
    \begin{multlined}
        \eval{\Omega^{(1)}_\text{GW, wall}(k)}_{H = \Gamma_\text{wall}} = 
        \frac{2 \pi \sigma^2}{3 \MPl^4 \Gamma_\text{wall}^{3/2} H_\text{re}^{1/2}} \\
        \times
        \begin{dcases}
            \qty(\frac{k}{\Gamma_\text{wall}})^3 \qty(\frac{\Gamma_\text{wall}}{H_\text{re}})^{3/2}, & k < \Gamma_\text{wall}, \\
            \qty(\frac{k}{H_\text{re}})^{3/2}, & \Gamma_\text{wall} \leq k < H_\text{re}, \\
            \frac{H_\text{re}}{k}, & k \geq H_\text{re},
        \end{dcases}
    \end{multlined}
\end{equation}
as shown in eq. (3.32) of Ref.~\cite{Bao:2024bws} in which $k = 2\pi f$ denotes the GW's wave-vector at $H = \Gamma_\text{wall}$, to be redshift to present-day frequencies. The abundance is further diluted by a factor of $\Omega_\text{rad}$ from the matter-radiation equality to today. Note that upon detecting the peak frequency and amplitude, $\sigma$ and $H_\text{re}$ can be inferred uniquely. 

The string spectrum is typically subdominant. As discussed in Ref.~\cite{Bao:2024bws}, the string part of the GW spectrum should be parametrically similar to that of a pure gauge string network that has been well studied previously \cite{Sousa:2020sxs, Sousa:2013aaa, Blanco-Pillado:2017oxo, Ringeval:2017eww, Cui:2017ufi, Cui:2018rwi, Vilenkin:1981iu, Hogan:1984is, Lorenz:2010sm, Auclair:2019wcv, Blanco-Pillado:2024aca, Siemens:2006yp, Olmez:2010bi, LIGOScientific:2013tfe, LIGOScientific:2017ikf}. This part of the spectrum $\Omega^{(1)}_\text{GW, str}(f)$ is almost flat with a UV roll-off governed by $H_\text{re}$ and an IR roll-off determined by $v_2$. Details are provided in \cref{app:GaugeStringSpectrum} of the Supplemental Material, and the final GW spectrum of the string-wall network is $\Omega^{(1)}_\text{GW} = \Omega^{(1)}_\text{GW, wall} + \Omega^{(1)}_\text{GW, str}$. 

\subsection{New Physics Scenario 2: Walls Bounded by Global Cosmic Strings \label{sec:globalModel}}
Previously, we discussed the GW spectrum from a string-wall system with boundary \emph{gauge} strings, but boundary strings can arise from breaking a \emph{global} symmetry. Then, boundary strings emit Nambu-Goldstone bosons (NGBs), opening up a new decay channel for the defect network. When the walls are large, the NGB emission from boundary strings is inefficient as the string comprises a subdominant component of the network. However, as walls shrink, the strings comprise a larger energy fraction of the network. Then, the NGB emission, with a rate $\propto k^2 \sim w^{-2}$, becomes efficient as the wall size $w$ decreases, modifying the UV part of the wall spectrum \cite{Bao:2024bws}. The scale at which the transition happens is $k_\text{NGB} \defeq \Gamma_\text{wall}^{1/2} \sigma^{1/2} / (60 v_2)$ with $v_2$ the VEV of the string-forming field. Physically, the NGB emission dominates when the wall size $w < k_\text{NGB}^{-1}$.

If $H_\text{re} < k_\text{NGB}$, the initial wall size is large such that the network primarily decays into GWs at first. The NGB emission becomes relevant only when the wall size is below $k_\text{NGB}^{-1}$ and sharpens the UV roll-off from $\propto f^{-1}$ to $\propto f^{-3}$, as shown in eq. (4.21) of Ref.~\cite{Bao:2024bws}
\begin{equation}\label{eq:scn2_main1}
    \begin{multlined}
        \eval{\Omega^{(2)}_\text{GW, wall}(k)}_{H = \Gamma_\text{wall}} = \Omega^{(1)}_\text{GW, wall}(k) \\
        \times
        \begin{dcases}
            1, & k < k_\text{NGB}, \\
            \qty(\frac{k_\text{NGB}}{k})^2, & k \geq k_\text{NGB}.
        \end{dcases}
    \end{multlined}
\end{equation}

If $H_\text{re} > k_\text{NGB}$ instead, the network can primarily decay into NGB immediately upon string re-entry. Once $H = \GammaNGBWall \defeq v_2^2 H_\text{re} / \sigma$ the typical NGB emission rate, the GW spectrum transitions from $\propto f^{3/2}$ growth to $\propto f^{-3}$ decay around $k \approx H_\text{re}$, which is given by
\begin{equation}\label{eq:scn2_main2}
    \begin{multlined}
    \eval{\Omega^{(2)}_\text{GW, wall}(k)}_{H = \GammaNGBWall}
    \approx \frac{2\pi \sigma^2}{3\MPl^4 (\GammaNGBWall)^{3/2} H_\text{re}^{1/2}} \\
    \times 
    \begin{dcases}
        \qty(\frac{k}{\GammaNGBWall})^3 \qty(\frac{\GammaNGBWall}{H_\text{re}})^{3/2}, & k \lesssim \GammaNGBWall, \\
        \qty(\frac{k}{H_\text{re}})^{3/2}, & \GammaNGBWall \lesssim k \lesssim H_\text{re}, \\
        \qty(\frac{H_\text{re}}{k})^3, & k \gtrsim H_\text{re}. 
    \end{dcases}
    \end{multlined}
\end{equation}

In both cases, a UV ``knee" frequency at which power laws change encodes the VEV $v_2$. Upon resolving the UV part of the GW spectrum, this may be used to infer the string tension without observing the string spectrum. 

The global string spectrum has been studied previously \cite{Figueroa:2020lvo, Chang:2021afa, Gorghetto:2021fsn, Baeza-Ballesteros:2023say}, and details are deferred to \cref{app:GlobalStringSpectrum} of Supplemental Material. Because of the efficient NGB emission, this signal is usually more suppressed, scaling as $\sim \mu^2/\MPl^4$ instead of $\sim \sqrt{\mu / \MPl^2}$ as in Scenario 1. Nonetheless, the total contribution is $\Omega^{(2)}_\text{GW} = \Omega^{(2)}_\text{GW, wall} + \Omega^{(2)}_\text{GW, str}$.

\section{Methodology \label{sec:methodology}}

Our analysis is a Bayesian framework that utilizes current available GW data and future projected sensitivities. Within a Bayesian framework, the posteriors for model parameters \(\vec{\theta}\) given data \(d\) (the set of pulsar timing residuals or $\hat{\Omega}_{\rm GW}$ estimates from Advanced LIGO) is $p(\vec{\theta} \mid d)
\;\propto\;
p(d \mid \vec{\theta})\,\pi(\vec{\theta})$
where \(\pi(\vec{\theta})\) is the prior and \(p(d\mid\vec{\theta})\) the likelihood. We conduct a search for a signal predicted by a cosmic domain wall and string network model. This signal is modeled through its contribution to the present-day gravitational-wave energy density spectrum, $\Omega_{\text{gw}}(f)$, which is defined by three primary parameters we denote as $\vec{\theta}=(\log_{10} H_{\rm re},\log_{10} v_2,\log_{10} \sigma)$. These parameters have different origin theoretically: $H_\text{re}$ from inflationary dynamics, $v_2$ from the first phase transition, and $\sigma$ from the second. We hence assign broad log-uniform priors to them without correlations: $\log_{10} H_{\rm re} (\rm GeV) \in \mathcal{U}(-34, 0)$, $\log_{10} v_2 (\rm GeV) \in \mathcal{U}(8, 16.5)$, and $\log_{10} \sigma (\rm GeV^3) \in \mathcal{U}(-5, 34)$. The upper bound of $H_\text{re}$ and $\sigma$, and the lower bound of $v_2$ are set by convenience. But the upper bound of $v_2$ is slightly above the GUT scale. The lower bound of $H_\text{re}$ is chosen mainly for convenience while ensuring no defects reentry on CMB scales, and the lower bound of $\sigma$ is to avoid low-tension domain wall formation, disrupting Big Bang Nucleosynthesis (BBN).

For the analysis of the NANOGrav 15-year dataset (NG15) \cite{NANOGrav:2023gor} we use the \textsc{ptarcade} \cite{Mitridate:2023oar} pipeline that is a high-level orchestration layer that builds and runs models in \textsc{ENTERPRISE}, which performs the core PTA likelihood valuations \citep{ellis_2020_4059815}. Data $d$ from NANOGrav's PTA measurement can be viewed as a vector of \emph{post-fit} timing residuals, $\delta\vec{t}$. It has been modeled $\delta\vec{t}$ as a zero-mean Gaussian process with total covariance $\mathbf{C}(\vec{\theta})$ determined by hyper-parameters $\vec{\theta}$ \citep{rasmussen2003gaussian,vanHaasteren:2012hj,van2014new}.

\captionsetup[figure]{skip=3pt}         
\captionsetup[subfigure]{skip=2pt,belowskip=0pt}  

\newlength{\TopBoxH} \setlength{\TopBoxH}{0.35\textheight} 
\newlength{\BotBoxH} \setlength{\BotBoxH}{0.25\textheight} 
\newcommand{\RowGap}{-0.25em} 

\afterpage{%
\begin{figure*}[!t]
  \centering

  \begin{subfigure}[b]{.5\textwidth}
    \centering
    \begin{minipage}[t][\TopBoxH][t]{\linewidth}
      \centering
      \includegraphics[
        width=1.50\linewidth,height=1.50\TopBoxH,keepaspectratio,
        trim=12 100 0mm 0,clip   
      ]{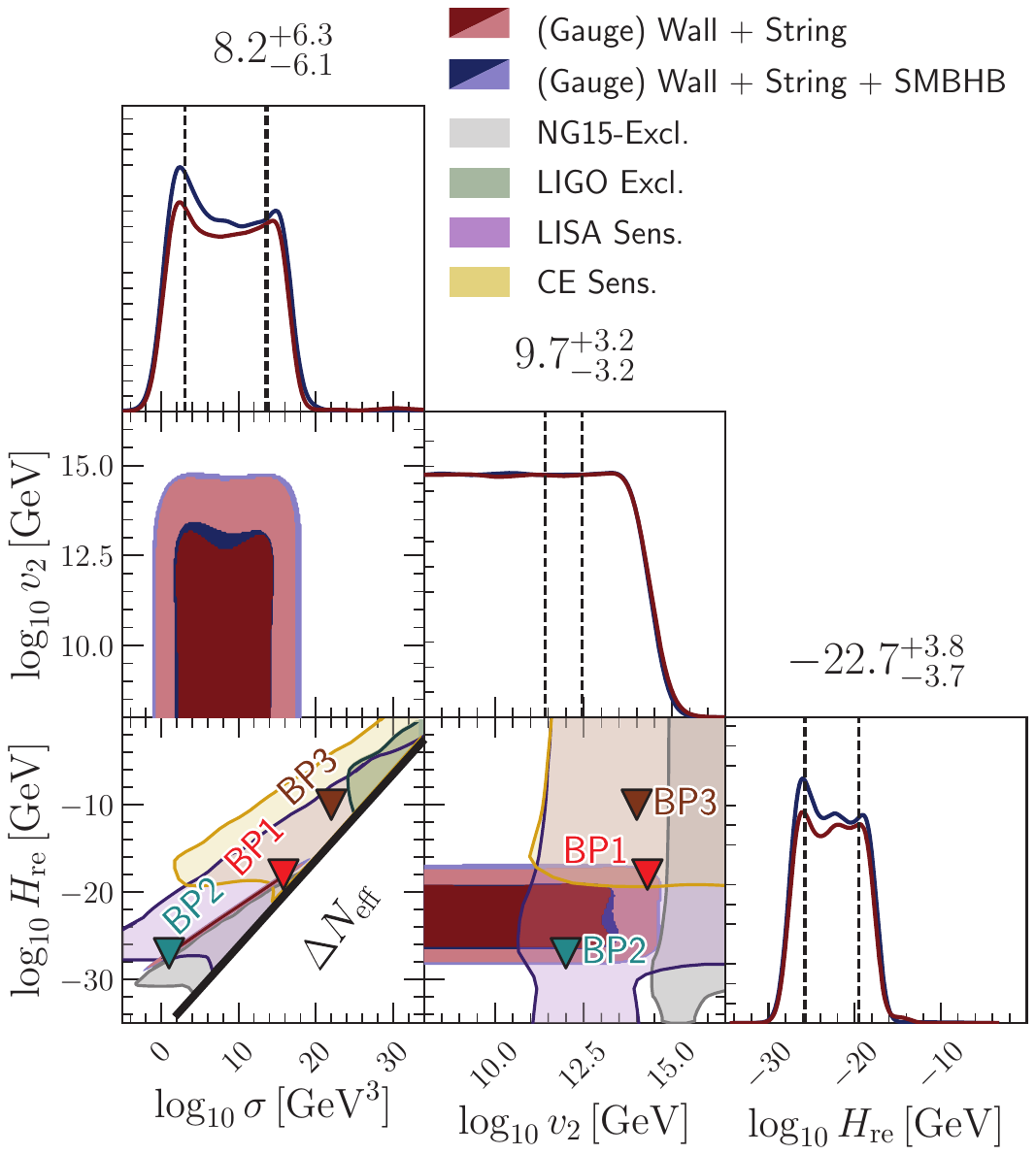}
    \end{minipage}
    \subcaption{}\label{fig:1a}
  \end{subfigure}
  \begin{subfigure}[b]{.5\textwidth}
    \centering
    \begin{minipage}[t][\TopBoxH][t]{\linewidth}
      \centering
      \includegraphics[
        width=1.25\linewidth,height=1.25\TopBoxH,keepaspectratio,
        trim=0.2mm 26mm 0mm 0mm,clip
      ]{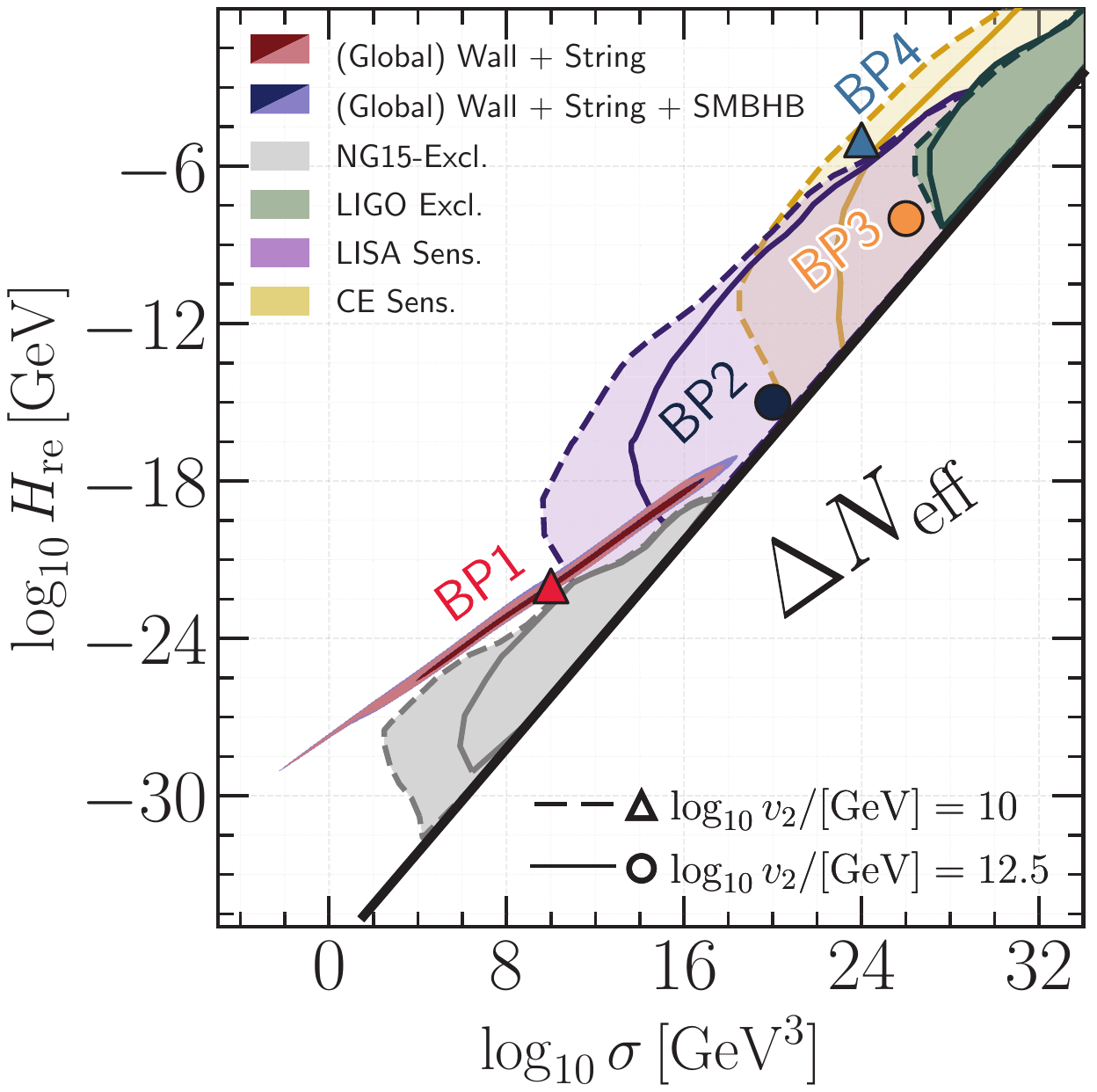}
    \end{minipage}
    \subcaption{}\label{fig:2a}
  \end{subfigure}
    
  \vspace{\RowGap} 

  \begin{subfigure}[b]{.5\textwidth}
    \centering
    \begin{minipage}[t]{\linewidth}
      \centering
      \includegraphics[
        width=1.\linewidth,height=\BotBoxH,keepaspectratio,
        trim=0 2mm 0 0,clip
      ]{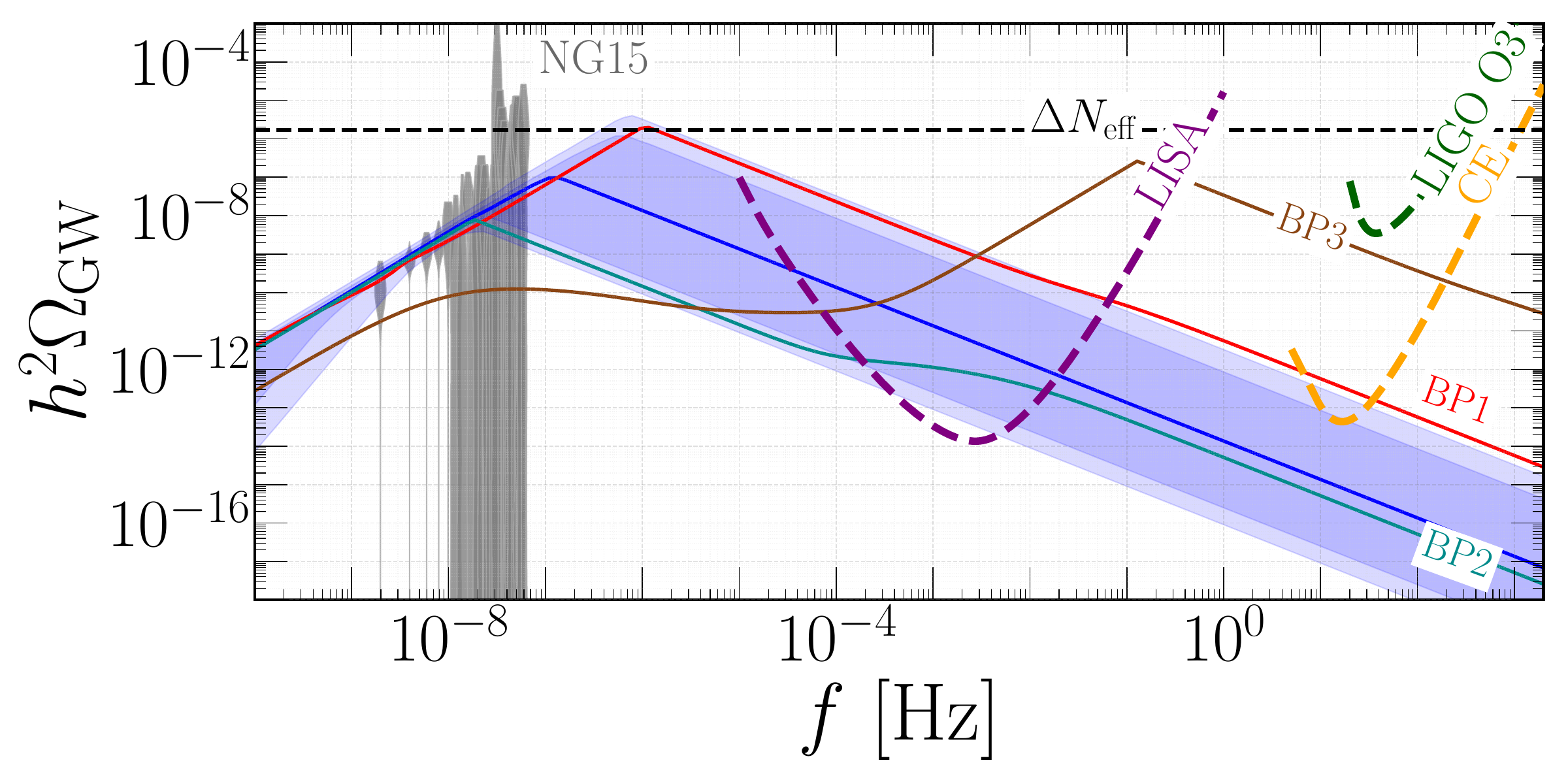}
    \end{minipage}
    \subcaption{}\label{fig:1b}
  \end{subfigure}
  \begin{subfigure}[b]{.5\textwidth}
    \centering
    \begin{minipage}[t]{\linewidth}
      \centering
      \includegraphics[
        width=1.\linewidth,height=\BotBoxH,keepaspectratio,
        trim=0 2mm 0 10,clip
      ]{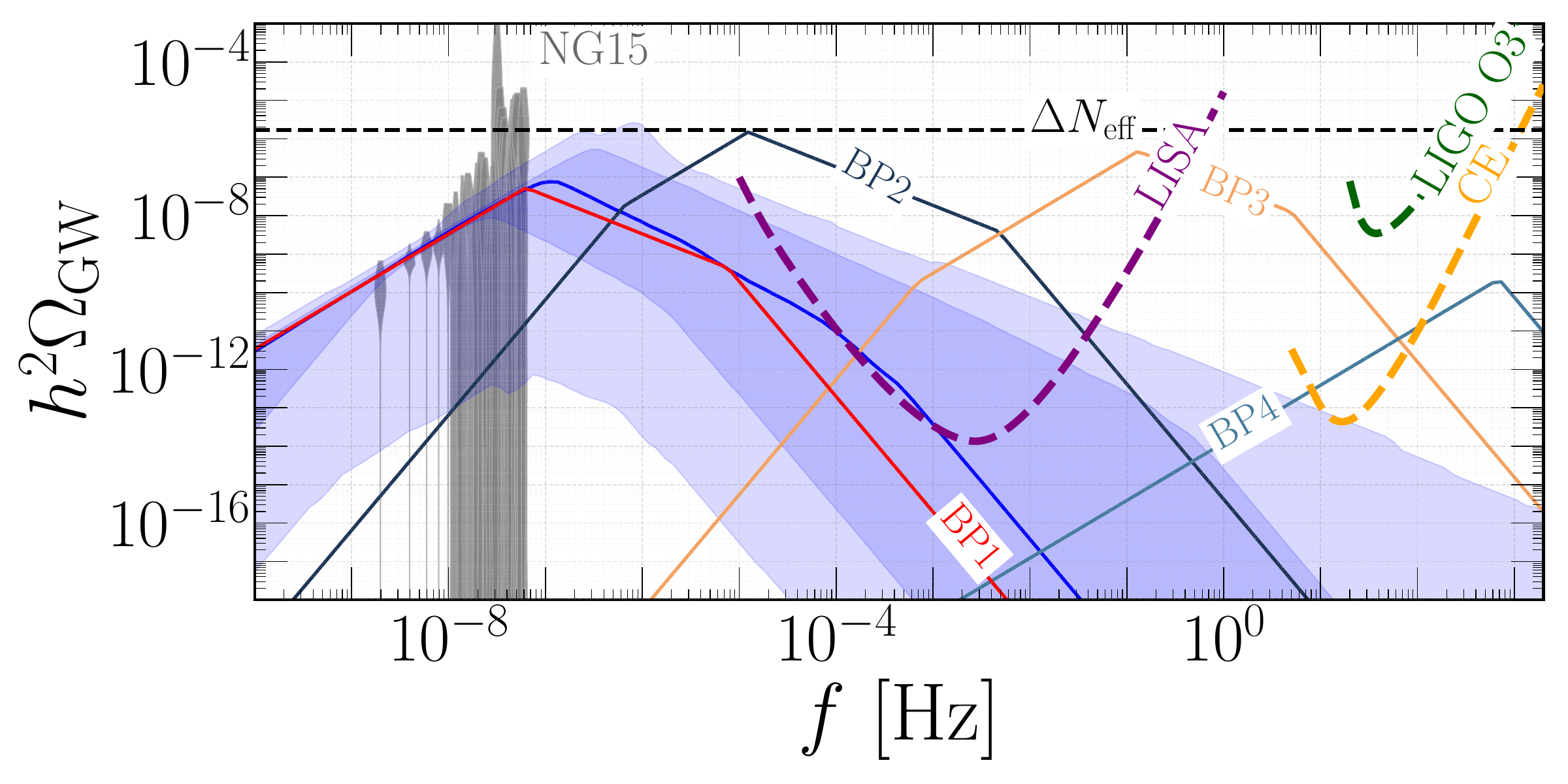}
    \end{minipage}
    \subcaption{}\label{fig:2b}
  \end{subfigure}

  \vspace{-0.35\baselineskip}
\caption{
\textbf{Left column:} Scenario~1 (walls bounded by \emph{gauge}
strings). \textbf{Right column:} Scenario~2 (walls bounded by \emph{global} strings).
\textbf{Conventions for upper panels:} dark (light) fill = 68\% (95\%) NANOGrav credible
regions. Overlays/colors follow the legend: \emph{red} = Wall+String;
\emph{blue} = Wall+String+SMBHB; \emph{light gray} = NG15 exclusion;
\emph{green} = LIGO exclusion; \emph{purple} = LISA sensitivity;
\emph{orange} = CE sensitivity. The diagonal \emph{black} line indicates
$\Delta N_{\rm eff}$ constraint. Benchmark points are labeled as BP$\#$.\textbf{ Upper panels:}
\textbf{(a)} Scenario~1: joint posteriors for
$(\log_{10}H_{\rm re},\,\log_{10}\sigma,\,\log_{10}v_2)$, marginalized over
SMBHB parameters.
\textbf{(b)} Scenario~2: \(\log_{10} H_{\rm re}\) vs.\ \(\log_{10}\sigma\) shown at two fixed values of \(\log_{10}(v_2/\mathrm{GeV})\): \(10\) and \(12.5\).
Dashed contours and triangular BP markers denote the \(10\) case, while solid contours and circular BP markers denote the \(12.5\) case. \textbf{Lower panels (spectra):}
\textbf{(c,d)} Posterior wall spectra for
the left (gauge) and right (global) scenarios. 
The light-blue band is the 95\% credible interval 
derived from the NANOGrav 15-year (NG15) data set 
and the solid blue curve the median of the wall spectra. (The string component enters
only as upper limits hence we do not include it in the posterior band.). We also show the spectra for corresponding benchmark points and detector sensitivities. See text for additional details and the full posteriors are shown in Supplemental Materials \cref{app:full_posteriors}. 
}
  
  \label{fig:unified_scenarios}
\end{figure*}
}

The likelihood for the PTA analysis is a multivariate normal,
\begin{equation}
p(\delta\vec{t}\mid\vec{\theta})
= 
\frac{\exp\!\big[-\tfrac12\,(\delta\vec{t})^{\mathrm{T}}\mathbf{C}^{-1}(\vec{\theta})\,\delta\vec{t}\big]}{\sqrt{(2\pi)^{N}\det \mathbf{C}(\vec{\theta})}},
\label{eq:likelihood}
\end{equation}
where $N$ is the number of residuals. The covariance decomposes as $
\mathbf{C}(\vec{\theta})=\mathbf{C}_{\mathrm{WN}}+\mathbf{C}_{\mathrm{RN}}+\mathbf{C}_{\mathrm{GWB}},
$ with $\mathbf{C}_{\mathrm{WN}}$ uncorrelated (white) measurement noise, $\mathbf{C}_{\mathrm{RN}}$ pulsar‑intrinsic
red noise, and $\mathbf{C}_{\mathrm{GWB}}$ the common red process from the SGWB. For an isotropic SGWB in GR,
inter‑pulsar correlations follow the Hellings–Downs overlap‑reduction function $\Gamma_{ab}$ \citep{hellings1983upper}. For this analysis, we also perform a model comparison against the standard SGWB signal attributed to a population of supermassive black hole binaries (SMBHBs), whose model and prior choices are detailed in \cref{app:S_SMBHB} of Supplemental Material. We also test a composite model containing both the domain wall-string and SMBHB signals. Model selection uses the Bayesian evidence $\mathcal{Z}$ and Bayes factor $K$:
\begin{equation}
\mathcal{Z}=\int p(\delta\vec{t}\mid\vec{\theta},\mathcal{M})\,p(\vec{\theta}\mid\mathcal{M})\,{\rm d}\vec{\theta},
\label{eq:evidence}
\end{equation}
\begin{equation}
K=\frac{\mathcal{Z}_1}{\mathcal{Z}_2}=\frac{p(\delta\vec{t}\mid\mathcal{M}_1)}{p(\delta\vec{t}\mid\mathcal{M}_2)}\,,
\label{eq:bayes_factor}
\end{equation}
where $K>1$ favors $\mathcal{M}_1$ over $\mathcal{M}_2$ \citep{kass1995bayes}.

For the interferometric detectors we use a Gaussian 
likelihood~\cite{allen1999detecting,flanagan1993sensitivity},
\begin{align}\label{bayesian}
\ln p(d\mid\vec{\theta})
&\propto
-\sum_k
\frac{\bigl[\hat{\Omega}(f_k) 
      - \Omega_{\text{GW}}(f_k\mid\vec{\theta})\bigr]^2}
     {\Omega^2_{\rm N}(f_k)}\,,
\end{align}
where $\hat{\Omega}(f_k)$ and $\Omega^2_{\rm N}(f_k)$ are 
the per-bin point estimates and variances. For Advanced LIGO, 
these are reported 
in~\cite{abbott2017upper,abbott2019search,abbott2021upper}. This expression is valid in the weak signal regime. 
For the LIGO O3 analysis no 
statistically significant SGWB has been detected, so this 
condition holds trivially. For CE and LISA, as shown in 
Fig.~S1, $\Omega_N(f) \gg \Omega_{\rm PLIS}(f)$ at all 
frequencies, so this condition is satisfied at the 
sensitivity contour boundaries.

For future detectors such as LISA and CE there is no realized dataset, so we present projected sensitivities.
Following the standard definition, sensitivity is characterized by the median significance for data generated under the null hypothesis $H_0$ \cite{Cowan:2010js}.
In our projections $H_0$ corresponds to the absence of any SGWB signal in the band, so the resulting sensitivity regions represent the best case, since additional astrophysical or instrumental contributions would only degrade the sensitivity.
Using the Gaussian likelihood in Eq.~(7) with the design sensitivity $\Omega_N(f_k)$ and the likelihood ratio $\Lambda(\vec{\theta})\equiv p(d\mid\vec{\theta})/p(d\mid\vec{0})$, one finds $\bigl\langle -2\ln\Lambda(\vec{\theta})\bigr\rangle_{H_0}=\sum_k \Omega_{\rm GW}^2(f_k\mid\vec{\theta})/\Omega_N^2(f_k)$, since $\langle\hat{\Omega}(f_k)\rangle_{H_0}=0$. The construction of the sensitivities $\Omega_{\rm N}(f_k)$ differ between the CE and LISA. For CE, we assume two detectors at the current LIGO sites and perform a cross-correlation measurement. For LISA, the three arms are coupled and therefore not independent; we assume the instrument noise is known and construct \(\Omega_{\rm N}^2(f_k)\) from auto-correlation measurements of the A and E TDI channels, as outlined in \cref{S_App1,S_App2} of the Supplemental Materials.

We also define the NANOGrav exclusion region using a one sided upper limit likelihood so that parameter values predicting spectra well below the data sensitivity are not excluded.
Using the NG15 per bin free spectrum with their kernel density representative posteriors, we set the likelihood as
$\ln p(d\mid\vec{\theta})=\sum_k \ln\!\left[\max_{b_k\ge 0}\ {\rm KDE}_k\!\big(\Omega_{\rm GW}(f_k\mid\vec{\theta})+b_k\big)\right]+\mathrm{const}$.

The resulting posteriors define the $95\%$ credible regions labeled “LIGO Excl.”, “LISA Sens.”, “CE Sens.”, and “NG15 Excl.”.

In addition to these, any proposed SGWB is subject to independent constraints from cosmology. By contributing to the total radiation energy density of the universe, a SGWB can alter large-scale observables, including the Cosmic Microwave Background (CMB) and the clustering of galaxies measured by Baryon Acoustic Oscillation (BAO) surveys \citep{Planck:2018vyg,BICEP:2021xfz}. The total energy density of a SGWB, integrated over all frequencies, is expressed as $h^2 \Omega_{\text{GW}} = \int_{0}^{\infty} h^2 \Omega_{\text{GW}}(f) d(\ln f)$.

This additional energy density effectively increases the number of relativistic species, $N_{\text{eff}}$. Observations from Big Bang Nucleosynthesis (BBN) and the CMB place firm upper limits on this value. For example, the Planck collaboration's analysis of the CMB power spectrum constrains the SGWB energy density to $h^2 \Omega_{\text{GW}} \lesssim 1.7 \times 10^{-6}$ at 95\% confidence \cite{Planck:2018vyg,Clarke:2020bil}. We exclude the regions violating this constraint in our analysis denoted by ``$\Delta N_{\rm eff}$".

\section{Results \label{sec:results}}
For both new physics scenarios, our MCMC analysis shows that several parameter regions are excluded at 95\% confidence level by current observations. The first is due to the $\Delta N_\text{eff}$ measurement jointly from CMB and BBN (triangular regions labeled $\Delta N_\text{eff}$ in \cref{fig:1a,fig:2a}). This bound applies to all cosmogenic SGWB and is challenging to surpass experimentally for high-frequency GW detectors. This is because current GW detectors measure strain while $\Delta N_\text{eff}$ constrains the energy density, which is proportional to $f^2$. At high frequencies, the constraints from $\Delta N_\text{eff}$ translate into strong bounds on the strain \cite{TitoDAgnolo:2024res}.

Fortunately, current GW detectors are reaching below $\Delta N_\text{eff}$, and near-future observatories will probe further. They contribute two more excluded regions. The NANOGrav excludes the gray regions in \cref{fig:1a,fig:2a}, and the LIGO O3 data excludes the green regions. For Scenario 1 (gauge-string scenario), the exclusion/sensitivity regions are shown in both the $H_\text{re}$-$\sigma$ and the $H_\text{re}$-$v_2$ plot. This is because $\Omega^{(1)}_\text{GW, wall}$ is independent of $v_2$, and $\Omega^{(1)}_\text{GW, str}$ of $\sigma$. Hence, the marginalized limits can be interpreted as limits on the wall part ($H_\text{re}$-$\sigma$ plot) and the string part ($H_\text{re}$-$v_2$ plot) of the GW spectrum, respectively. Then, the gray region on the $H_\text{re}$-$v_2$ plane shows that NANOGrav excludes $v_2 \gtrsim \SI{E15}{\GeV}$ (or $G\mu \gtrsim 2\times 10^{-8}$) for the string spectrum, while the current LIGO constraint on $v_2$ remains weaker (outside the plotted range). However, the wall part is excluded by both NANOGrav and LIGO at different corners of \cref{fig:1a}, demonstrating the advantage of multi-band GW searches. We do not show additional contours in $v_2$-$\sigma$ plane for \cref{fig:1a} because the marginalized parameter, $H_\text{re}$, alters both string and wall parts, and a marginalized exclusion region on the $v_2$-$\sigma$ plane is strongly prior-dependent without a clean physical interpretation. 

Similarly, for Scenario 2 (global-string scenario), the GW spectrum depends on all three parameters jointly. Thus, we report the exclusion contours on $H_\text{re}$-$\sigma$ plane for fixed $v_2$. While the contours change with $v_2$, both scenarios exclude similar corners in the $H_\text{re}$-$\sigma$ plane, showing consistency. 

\begin{table}[t]
  \centering
  \large
  \makebox[\columnwidth][c]{%
    \begin{tabular}{c | *{2}{c}}
      \hline\hline
           & Network & Network + SMBHB \\
      \hline
      Gauge  & $17.3 \pm 1.5$ & $22.2 \pm 1.8$ \\
      Global & $0.37 \pm 0.06$& $6.3 \pm 0.6$  \\
      \hline\hline
    \end{tabular}%
  }%
  \begingroup\makeatletter
    \let\@capwidth\columnwidth
  \makeatother
  \caption{Bayes factors of interpretations of NANOGrav data: This is
  normalized to $1$ for SMBHB-only foreground. A larger Bayes factor implies the new physics interpretation is preferred over the SMBHB-only interpretation.}
  \label{tab:bayesFactors}
  \endgroup
\end{table}

Beyond the current exclusions, one may attribute the observed NANOGrav signal to this string-wall network. The best-fit regions to NANOGrav data corresponds to fits with (blue in \cref{fig:1a,fig:2a}) and without (red) the SMBHB foreground's prior.%
\footnote{We give the full set of posteriors in Supplemental Materials in \cref{fig:triangle_two_chains_gauge,fig:triangle_two_chains_global,fig:post_1,fig:post_1_close,fig:post_1_global,fig:post_2,fig:post_2_close,fig:post_2_close_global,fig:post_1_close_global,fig:post_2_global}.}
The two regions' strong overlap shows compatibility between the two interpretations. Bayes factor analysis also delivers a similar message. As seen in \cref{tab:bayesFactors}, the Bayes factors increase as we include the SMBHB foreground for both scenarios. A higher Bayes factor implies a string-wall interpretation is preferred over the standard SMBHB interpretation, and except for one case, all Bayes factors favor ($>1$) new physics but not decisively ($>10^2$ required). The Bayes factor is small for the global-string scenario without SMBHB; however, \cref{fig:2b} shows a reasonable fit to NANOGrav data. This is because Bayes factors automatically penalize model complexity by lowering the marginalized likelihood for non-explanatory parameter space. Here, the $v_2$ dependence in $\Omega_\text{GW, wall}^{(2)}$ only modifies the UV roll-off of the wall spectrum and is not measured by NANOGrav. Hence, $v_2$ is mostly a nuisance parameter to explain NANOGrav data and is disfavored in a Bayes factor analysis. This, however, can be an advantage as we discuss next. 

We also provide $95\%$ projected sensitivities of future detectors, such as LISA and CE. Combining the NANOGrav fit with future reaches yields particularly interesting conclusions. For Scenario 1, LISA sensitivity (purple) covers almost entirely the best-fit region in the $H_\text{re}$-$\sigma$ plane of \cref{fig:1a}. Thus, LISA mission can decisively test the string-wall interpretation, because the best-fit spectrum of Scenario 1 (blue curve and band of \cref{fig:1b}) has UV tails within LISA's sensitivity. CE (orange) carves a smaller best-fit region, but Benchmark Point (BP) 1 of \cref{fig:1a,fig:1b} illustrates a parameter combination with a CE-detectable signal. While the CE sensitivity seems irrelevant for NANOGrav, a joint signal in LISA and CE implies that $H_\text{re} \gtrsim \SI{E-20}{\GeV}$ as shown in \cref{fig:1a}. Thus, complementary GW searches across frequency bands are essential. If LISA observes a partially flat spectrum (BP2 of \cref{fig:1a,fig:1b}), then the gauge-string interpretation with $v_2 \gtrsim \SI{E11}{\GeV}$ is strongly favored as the $H_\text{re}$-$v_2$ plot of \cref{fig:1a} shows. 

For Scenario 2, LISA probes only part of NANOGrav's fits. Some NANOGrav-fitting signal even evades LISA's sensitivity (BP1 of \cref{fig:2a,fig:2b}). Because $v_2$ controls the UV knee (\cref{eq:scn2_main1}), LISA and NANOGrav may jointly map the peak frequency, peak amplitude, and knee frequency, thus determining all three model parameters and inferring the string tension even without detecting the string component. CE measurement is also beneficial because it is less likely for NANOGrav fits to be observed by CE in Scenario 2 than in Scenario 1 (comparing \cref{fig:1b} with \cref{fig:2b}). Thus, a positive CE detection favors Scenario 1 over Scenario 2. Lastly, it is worth commenting that our NANOGrav best-fit region peaks in the microhertz range, a frequency band under active research. Proposals using interferometry, photometric surveys, satellite and lunar laser ranging, fast radio burst timing, orbiter Doppler tracking, \textit{etc.}, show potential of probing SGWB in this frequency band despite its challenge \cite{Sesana:2019vho, Wang:2020pmf, Blas:2021mqw, Caliskan:2023cqm, Zwick:2024hag, Fedderke:2021kuy, Lu:2024yuo, Foster:2025nzf}.

Even if the NANOGrav signal has another origin, these models predict broad, correlated features across frequencies, calling for multi-band GW detection. For instance, BP3s of both Scenario 1 and 2 show how a correlated signal shows up in LISA-CE joint detection. LISA distinguishes the two scenarios by detecting a flat IR part against a roll-off. CE, then, can potentially observe the UV knee if Scenario 2 is realized, thereby determining $v_2$ synergistically. A positive detection in one detector accompanied with a null detection in the other is also informing as illustrated by BP2 and BP4 of \cref{fig:2a,fig:2b}. BP2's peak and knee can be resolved by LISA, allowing all three model parameters to be fitted. Then, the negative detection in CE corroborates the string-wall interpretation of the GW signal. If CE observes another peak after foreground subtraction, then the naive model considered here does not explain the SGWB, although simple adjustments to the model, such as adding additional defects, may accommodate it. BP4 illustrates a case when CE makes an observation while LISA remains silent. Because fewer spectral features can be identified by CE for BP4, determining the precise model parameters is challenging. Nonetheless, a high $H_\text{re}$ and $\sigma$ is preferred as \cref{fig:2a} shows, which cannot be inferred without LISA's null detection.

\section{Conclusion and Discussion \label{sec:conclusion}}
This paper highlights the multi-band nature of current and near-future GW astrophysics using a new physics scenario as a testbed. The model has two phase transitions: the first generates boundary cosmic strings, the second domain walls. An inflationary epoch occurs between the two phase transitions, stretching the strings outside the horizon, letting domain walls dominate the network and generate large GW signals. While the parametric argument leaves some theoretical uncertainties on the model parameters, we expect the spectral features to remain qualitatively unchanged. Current detectors (LVK and NANOGrav) probe two frequency bands with a positive detection in the nHz range. Future detectors, such as CE, ET, and LISA, will expand the coverage with higher sensitivity. 

We offered a uniform pipeline incorporating the current limits, the observed signal, and the future sensitivities into a multi-band analysis. Current observations can exclude scattered corners of the parameter space, leaving ample room to explore. The new physics scenario can explain the observed NANOGrav background, both without and with the expected astrophysical foreground, but not decisively. Future observations on the UV roll-off can test this interpretation, underscoring the importance of multi-band analysis in mapping out the GW landscape.

We also note that this multi-band analysis has further room for improvement. On the theory frontier, the uncertainties in the spectrum invite improvements by future detailed studies. Given the complex dynamics of defects, an efficient and reliable numerical method to determine $\Omega_\text{GW}(f | \vec{\theta})$ and its uncertainty can improve our current analysis. On the observation frontier, the expected astrophysical foreground is understood crudely and needs better modeling informed by GW observations. Incorporating a simulated foreground into the pipeline will further enable searches for new physics signatures atop astrophysical backgrounds.

\acknowledgments
We thank Huai-ke Guo for the helpful comments on the draft. YB and LTW are supported by the Department of Energy grant DE-SC0013642. TB and VM are in part supported by the NSF grant PHY-2409173. This document has a LIGO document number of P2500693.

\appendix

\bibliography{SGWB_Across_Frequency}

\clearpage
\pagebreak
\widetext
\begin{center}
\textbf{\large Supplemental Materials: Searching Stochastic Gravitational Wave Background Landscape Across Frequency
Bands}
\end{center}
\setcounter{equation}{0}
\setcounter{figure}{0}
\setcounter{table}{0}
\setcounter{page}{1}
\makeatletter
\renewcommand{\theequation}{S\arabic{equation}}
\renewcommand{\thefigure}{S\arabic{figure}}
\renewcommand{\bibnumfmt}[1]{[#1]}
\renewcommand{\citenumfont}[1]{#1}
\setcounter{section}{0}
\renewcommand{\thesection}{S\Roman{section}}

\section{Construction of SGWB Sensitivities}\label{S_App1}
The metric perturbation in the frequency domain decomposed into two ($+$ and $\times$) polarizations is \cite{maggiore2008gravitational}
\begin{equation}\label{strain1}
h_{ab}(f,\hat{\Omega})
= h_+(f,\hat{\Omega})\,e^+_{ab}(\hat{\Omega}) + h_{\times}(f,\hat{\Omega})\,e^{\times}_{ab}(\hat{\Omega})
= \sum_{A=+,\times} e^A_{ab}(\hat{\Omega})\,h_A(f,\hat{\Omega})\,.
\end{equation}

\begin{align}
e^{+}_{ab}(\hat{\Omega}) &= \hat{k}_a \hat{k}_b - \hat{l}_a \hat{l}_b \label{eq:polbasis1},\\
e^{\times}_{ab}(\hat{\Omega}) &= \hat{k}_a \hat{l}_b + \hat{l}_a \hat{k}_b \label{eq:polbasis2}, 
\end{align}

\noindent where
\begin{align}
\hat{\Omega} &= \cos\phi\,\sin\theta\, \hat{x} + \sin\phi\,\sin\theta\, \hat{y} + \cos\theta\, \hat{z} , \\
\hat{l} &= \cos\phi\,\cos\theta\, \hat{x} + \sin\phi\,\cos\theta\, \hat{y} - \sin\theta\, \hat{z} ,\\
\hat{k} &= \sin\phi\, \hat{x} - \cos\phi\, \hat{y} , 
\end{align}
For an isotropic, unpolarized background one has \cite{allen1999detecting}
\begin{equation}\label{eq:polcor1}
\bigl\langle h_A(f,\hat{\Omega})\,h_{A'}^*(f',\hat{\Omega}')\bigr\rangle
= \frac{1}{8\pi}\,S_h(f)\,\delta(f-f')\,\delta_{AA'}\,\delta^2(\hat{\Omega},\hat{\Omega}')\,,
\end{equation}
with \(A,A'\in\{+,\times\}\).  Hence
\begin{align}\label{eq:polcor2}
\bigl\langle h_{ab}(f,\hat{\Omega})\,h^{*}_{cd}(f',\hat{\Omega}')\bigr\rangle
&= \sum_{A,A'} e^A_{ab}(\hat{\Omega})\,e^{A'}_{cd}(\hat{\Omega}')
   \bigl\langle h_A(f,\hat{\Omega})\,h_{A'}^*(f',\hat{\Omega}')\bigr\rangle \nonumber\\
&= \frac{1}{8\pi}\,S_h(f)\,\delta(f-f')\,\delta^2(\hat{\Omega},\hat{\Omega}')
   \sum_{A} e^A_{ab}(\hat{\Omega})\,e^{A}_{cd}(\hat{\Omega})\nonumber\\
&= \frac{1}{8\pi}\,S_h(f)\,\delta(f-f')\,\delta^2(\hat{\Omega},\hat{\Omega}')\,
   P_{abcd}(\hat{\Omega})\,,
\end{align}
where
\begin{align}
P_{abcd}(\hat{\Omega})
= \frac{1}{2}\bigl(P_{ac}P_{bd} + P_{ad}P_{bc} - P_{ab}P_{cd}\bigr),
\quad
P_{ab} = \eta_{ab} - n_a n_b.
\end{align}
Energy density of gravitational waves is modeled in terms of the dimensionless quantity
\begin{align}
    \Omega_{\rm GW} (f) = \frac{1}{\rho_c} \frac{d\rho_{\rm GW}}{d\ln f}
\end{align}
with $\rho_c = 3H^2_0 c^2/(8\pi G)$ and
\begin{align}
    \rho_{\rm GW} = \frac{c^2}{32 \pi G} \braket{\dot{h}_{ab} \dot{h}^{ab} }
\end{align}
Using \cref{strain1,eq:polcor1,eq:polcor2} with $\frac{1}{4}\sum_{A=+,\times}  e^A_{ab}(\hat n)  e_{ab,A}(\hat n)=1$, one can show that
\begin{align}
    \Omega_{\rm GW}(f) = \frac{4\pi^2f^3}{3 H^2_0} S_h(f) .
\end{align}

Following \cite{Romano:2016dpx}, the strain measured by an ideal (noise-free) detector at time \(t\)  
\begin{equation}
h(t)
= \int_{-\infty}^{\infty}df\ e^{i2\pi f t}\,
  \int_{S^2}d\Omega\;
  R^{ab}(f,\hat{\Omega})\,h_{ab}(f,\hat{\Omega}),
\end{equation}
where \(R^{ab}(f,\hat{\Omega})\) is the detector response tensor.  For  Michelson interferometer pairs with two arms with arm–unit vectors \(\mathbf{u},\mathbf{v}\) and vertex at \(\mathbf{x}_0\) \cite{Romano:2016dpx},
\begin{equation}\label{transfer}
R^{ab}(f,\hat{\Omega})
= \frac{1}{2}\,e^{2\pi i f\,\mathbf{x}_0\cdot\hat{\Omega}/c}
  \Bigl[u^a u^b\,T\bigl(f,\hat u\!\cdot\!\hat{\Omega}\bigr)
       - v^a v^b\,T\bigl(f,\hat v\!\cdot\!\hat{\Omega}\bigr)\Bigr],
\end{equation}
with the single-arm transfer function
\begin{equation}
T(f, \hat u \cdot \hat{\Omega}) =  e^{i f / f_*} \left[ e^{-i \frac{f}{2f_*}(1 - \hat u \cdot \hat{\Omega})} \mathrm{sinc}\left(\frac{f}{2f_*}[1 + \hat u \cdot \hat{\Omega}]\right) + e^{i \frac{f}{2f_*}(1 + \hat u \cdot \hat{\Omega})} \mathrm{sinc}\left(\frac{f}{2f_*}[1 - \hat u \cdot \hat{\Omega}]\right) \right]
\end{equation}
and the characteristic frequency
\begin{equation}
f_* = \frac{c}{2\pi L}\,.
\end{equation}
For \(f\ll f_*\) (the small‐antenna limit) one has \(T\simeq1\).  In the case of LISA (\(L\sim2.5\times10^9\) m, \(f_*\simeq 2\) mHz) and CE (\(L\sim4.0\times10^4\) m, \(f_*\simeq1.2\) kHz) the full frequency dependence of \(T\bigl(f,\hat u\!\cdot\!\hat{\Omega}\bigr)\) must be retained, which suppresses sensitivity above \(f_*\). For the case of LIGO and CE , $f_* \sim 12$ kHz \cite{LIGOScientific:2014pky} and the most sensitive band is well below $f_*$ so one can set $T(f, \hat u \cdot \hat{\Omega})\simeq1$.

We assume an additive Gaussian noise ($n_I(t)$) besides the signal
\begin{equation}
    s_I (t) = h_I(t) + n_I(t)
\end{equation}
where $I=1,2\dots N$ denotes the detector index.

The cross‐correlation statistic for two detectors is
\begin{equation}
Y = \int_{-\infty}^{\infty}df\ \;
    \tilde s_1(f)\,\tilde s_2^*(f)\,\tilde Q(f)\,,
\label{eq:cross_corr_freq}
\end{equation}
where \(\tilde Q(f)\) is a filter to be determined.  The signal‐to‐noise ratio is
\begin{equation}
\mathrm{SNR} = \frac{\langle Y\rangle}{\sigma_Y}\,.
\end{equation}

Assuming uncorrelated detector noise and a stationary SGWB,
\begin{equation}
\langle Y\rangle
= \int df\ \;\langle \tilde h_1(f)\,\tilde h_2^*(f)\rangle\,\tilde Q(f)
= \frac{T}{2}\int_{-\infty}^{\infty} df \;S_h(f)\,\gamma_{12}(f)\,\tilde Q(f),
\label{eq:Y_mean}
\end{equation}
where
\begin{equation}
\gamma_{12}(f)
= \frac{5}{8\pi}\sum_{A=+,\times}
  \int_{S^2} d\Omega\;
  e^{-2\pi i f\,\hat{\Omega}\cdot\Delta\mathbf{x}/c}\,
  F_1^A(\hat{\Omega},f)\,F_2^A(\hat{\Omega},f)
\label{eq:orf}
\end{equation}
is the \emph{overlap reduction function} \cite{flanagan1993sensitivity,allen1999detecting}, with \(\Delta\mathbf{x}=\mathbf{x}_1-\mathbf{x}_2\) and
\(F_i^A(\hat{\Omega},f)=e^A_{ab}(\hat{\Omega})R_i^{ab}(f,\hat{\Omega})\).  A conventionally chosen prefactor \(5/(8\pi)\) is applied so that \(\gamma_{12}(0)=1\) for co‐aligned, co‐located perpendicular arm detectors.

In the weak‐signal limit (\(h\ll n\)) and assuming Gaussian, stationary noise \cite{allen1999detecting}
\begin{equation}
\sigma_Y^2
= \frac{T}{4}\int_{-\infty}^{\infty} d f\;P_1(f)\,P_2(f)\,\bigl|\tilde Q(f)\bigr|^2,
\label{eq:Y_var}
\end{equation}
where the one‐sided noise PSD is defined by
\(\langle n_I^*(f)n_J(f')\rangle = \tfrac12\,\delta(f-f')\delta_{IJ}\,P_I(f)\).

We define the inner product
\begin{equation}
(A,B)
\;=\;\int_{-\infty}^{\infty}df\;
  A^*(f)\,B(f)\,P_1(f)\,P_2(f).
\end{equation}

Then using \eqref{eq:Y_mean} and \eqref{eq:Y_var},
\begin{align}
\mathrm{SNR}^2
= \frac{\langle Y\rangle^2}{\sigma_Y^2}
=\frac{(\frac{3H_0^2\,T}{20\pi^2})^2\,
  \biggl(\tilde Q,\;\frac{\Omega_{\rm GW}(f)\,\gamma_{12}(f)}{f^3\,P_1\,P_2}\biggr)^2}{\frac{T}{4}\,(\tilde Q,\tilde Q)} =  \Bigl(\tfrac{3H_0^2}{10\pi^2}\Bigr)^2\,T\,
  \frac{\bigl(\tilde Q,\frac{\Omega_{\rm GW}\,\gamma_{12}}{f^3\,P_1\,P_2}\bigr)^2}
       {(\tilde Q,\tilde Q)}\,.
\end{align}
Maximizing SNR gives the well‐known matched filter
\begin{equation}
\tilde Q(f)
= \lambda\,\frac{\Omega_{\rm GW}(f)\,\gamma_{12}(f)}{f^3\,P_1(f)\,P_2(f)}\,,
\label{eq:opt_filter}
\end{equation}
with normalization \(\lambda\) which cancels out from the final SNR result. One finds for a narrow‐bin width \(\delta f\)
\begin{equation}\label{sigmaf}
\sigma(f)
= \frac{10\pi^2}{3H_0^2}\,
  \frac{f^3\sqrt{P_1(f)\,P_2(f)}}{|\gamma_{12}(f)|}\,
  \sqrt{\frac{1}{2\,T\,\delta f}}\,.
\end{equation}

We can define effective noise $\Omega_{\rm N}(f)$ expressed in terms of a GW energy density spectrum as 
\begin{align}\label{omega_n}
    \Omega^{}_{\rm N}(f) = \frac{10\pi^2}{3H_0^2}\,
  \frac{f^3\sqrt{P_1(f)\,P_2(f)}}{|\gamma_{12}(f)|} 
\end{align}

This expression is suitable for LIGO and CE where the noise spectral densities refer to Hanford/Livingston detectors and the proposed CE design sensitivity respectively. For multiple baselines one uses $\Omega_N(f) = \Big[\sum_{I<J} \sum_J (\Omega^{IJ}_{N}(f))^{-2} \Big]^{-1/2}$.

Based on this SNR can be expressed as:
\begin{align}
    \rm SNR = \Big[T \int^{\infty}_{0} df \Big( \frac{\Omega_{\rm GW}(f)}{\Omega_{\rm N}(f)} \Big)^2\Big]^{1/2}
\end{align}

The construction above assumes two independent detectors with uncorrelated noise. LISA is a single constellation. LISA shape can be shown to be algebraically equivalent, for an isotropic SGWB, to two co-located Michelson interferometers each with opening angle \(90^\circ\), whose principal arms are rotated by \(\pm45^\circ\) relative to each other. We have three single arm interferometers based at each vertices, $\alpha=X,Y,Z$ \cite{Prince:2002hp}. We can follow the previous discussion and construct three $V$ shaped interferometers separated by $L$. However, our assumption that noise is uncorrelated in each detector does not hold anymore since each pair shares an arm. Therefore, our noise correlation matrix takes the form \cite{Prince:2002hp},
\begin{align}
    \braket{n_{\alpha} (f) n_{\alpha'}(f')} = \frac{1}{2} \delta(f-f') \begin{pmatrix}
     N_d & N_o & N_o\\
     N_o & N_d & N_o\\
     N_o & N_o & N_d
    \end{pmatrix}(f)
\end{align}
where we assumed each detector to be identical. It is customary to define the following data strain channels $A,E,T$  \cite{cornish2001space}
\begin{align}
    &s_A = \frac{s_Z-s_X}{\sqrt{2}} \\
    &s_E = \frac{s_X-2s_Y+s_Z}{\sqrt{6}}\\
    &s_T = \frac{s_X+s_Y+s_Z}{\sqrt{3}}.
\end{align}
so that
\begin{align}
\bigl\langle s_T(f)\,s_T^*(f')\bigr\rangle &= \frac{1}{2}\,\delta(f - f')\bigl[N_d(f) + 2\,N_o(f)\bigr]\,,\\
\bigl\langle s_A(f)\,s_A^*(f')\bigr\rangle &= \frac{1}{2}\,\delta(f - f')\bigl[N_d(f) - N_o(f) + \mathcal{R}_{I}(f)S_h(f)\bigr]\,,\\
\bigl\langle s_E(f)\,s_E^*(f')\bigr\rangle &= \frac{1}{2}\,\delta(f - f')\bigl[N_d(f) - N_o(f) +\mathcal{R}_{I}(f) S_h(f)\bigr]\,,\\
\bigl\langle s_T(f)\,s_A^*(f')\bigr\rangle 
  &= \bigl\langle s_A(f)\,s_T^*(f')\bigr\rangle = 0\,,\\
\bigl\langle s_T(f)\,s_E^*(f')\bigr\rangle 
  &= \bigl\langle s_E(f)\,s_T^*(f')\bigr\rangle = 0\,,\\
\bigl\langle s_E(f)\,s_A^*(f')\bigr\rangle 
  &= \bigl\langle s_A(f)\,s_E^*(f')\bigr\rangle 
     = 0\,.
\end{align}

$\mathcal{R}_{I}(f)$ is the sky-averaged detector response to the gravitational wave for channel $I$,  related to the previously defined antenna functions $F^A_I(\hat{\Omega},f)$ for the channel as, 
\begin{equation}
\mathcal{R}_{I}(f) = \sum_{A = +, \times} \int \frac{d \Omega}{4 \pi} \, | F^A_{I} (\hat{\Omega}, f) |^2.
\end{equation}
Therefore, one identifies two independent A and E channels (unit auto-normalization, $\gamma_{AA}=\gamma_{EE}=1$) with \emph{vanishing} cross-correlation for an isotropic background ($\gamma_{AE}=0$); T is a null channel \cite{cornish2001space}. Therefore the two‑detector
cross‑correlation statistic does not apply to LISA; instead one should use the
\(A\) and \(E\) \emph{auto‑powers} with noise subtraction.

Assuming perfect knowledge of \(P_{n,I}(f) = N_d(f)-N_o(f)\) so that the noise bias in \(|s_I|^2\) can be
removed, the optimal quadratic estimator built from \(A\) and \(E\) (matched weights in the
weak‑signal limit) gives
\begin{align}
\mathrm{SNR}^2
= T\sum_{I=A,E}\int_0^\infty df\;
\frac{\,[\mathcal{R}_{I}(f)S_h(f)]^2}{P_{n,I}(f)^2}.
\end{align}
It is convenient to quote per‑channel curves \(\Omega_{N,I}(f)\) satisfying
\begin{align}
\Omega_{N,I}(f)
=\frac{2\pi^2}{3H_0^2}\,
\frac{f^{3}\,P_{n,I}(f)}{\sqrt{T}\mathcal{R}_{I}(f)},
\qquad
\frac{1}{\bigl[\Omega_N^{\rm LISA}(f)\bigr]^2}
=\frac{1}{\Omega_{N,A}(f)^2}+\frac{1}{\Omega_{N,E}(f)^2} \simeq \frac{2}{\Omega_{N,A}(f)^2}
\end{align}
In the final step we assumed identical noise for both channels.

\section{Noise Power Spectral Densities}\label{S_App2}
\label{sec:psd}

Cosmic Explorer comprises two L-shaped detectors in the United States which we assumed to be at current Hanford/Livingston LIGO detector locations, each with arm length $L_{\rm CE}=40\,\mathrm{km}$ \cite{punturo2010third}. The design noise sensitivities can be found in \cite{LIGOScientific:2016wof}.

LISA is a heliocentric, equilateral-triangle constellation with armlength $L_{\rm LISA}=2.5\times10^9\,\mathrm{m}.$ \cite{amaro2017laser}.

The noise spectrum for LISA can be expressed as the sum of two separate components \cite{amaro2017laser,Robson:2018ifk}:
\begin{equation}
P_{\text{noise}}^{\text{LISA}}(f) = \frac{1}{(L_\text{LISA})^2} \left[ P_{\text{oms}}^{\text{LISA}}(f) + \frac{2}{(2\pi f)^4} \left( 1 + \cos^2\left(\frac{f}{f_*^{\text{LISA}}}\right) \right) P_{\text{acc}}^{\text{LISA}}(f) \right],
\tag{A.41}
\end{equation}

In this equation, the terms $L_\text{LISA}$ (LISA's arm length) and $f_*^{\text{LISA}}$ (characteristic frequency) were mentioned previously. The terms $P_{\text{oms}}^{\text{LISA}}$ and $P_{\text{acc}}^{\text{LISA}}$ correspond to the position noise from the \textit{optical metrology system} (OMS) and the acceleration noise from a single test mass, respectively.
\begin{align}
P_{\text{oms}}^{\text{LISA}}(f) &\simeq (1.5 \times 10^{-11} \, \text{m})^2 \left[ 1 + \left(\frac{2\,\text{mHz}}{f}\right)^4 \right] \text{Hz}^{-1}, \tag{A.42} \\
P_{\text{acc}}^{\text{LISA}}(f) &\simeq (3 \times 10^{-15} \, \text{m}\,\text{s}^{-2})^2 \left[ 1 + \left(\frac{0.4\,\text{mHz}}{f}\right)^2 \right] \left[ 1 + \left(\frac{f}{8\,\text{mHz}}\right)^4 \right] \text{Hz}^{-1}. \nonumber
\end{align}

In addition to these instrumental contributions, LISA will see an unresolved “confusion” foreground—primarily from Galactic white‐dwarf binaries—that raises the low–frequency noise floor. In practice, much of this confusion noise can be mitigated leveraging the anisotropies in LISA and  astrophysically motivated templates \cite{Criswell:2024hfn}. For simplicity, however, we will ignore any residual confusion and treat $P_n^{\rm LISA}(f)$ as purely instrumental and known.

For each detector we assume $T=3$ yr observation time. We use these to construct $\Omega_{\rm N}(f)$ for our Bayesian search for each detector network in Eq. \eqref{bayesian}. Although not used in our analysis we also construct the so called Power Law Integrated Sensitivity (PLIS) curves \cite{Thrane:2013oya} for the previously mentioned interferometric detectors to roughly judge the detectability of our analysis posteriors.

For this construction the signal's amplitude, $\Omega_{\text{signal}}(f)$, is modeled as a power law:
\begin{equation}\label{eq:signaleq}
    \Omega_{\text{signal}}(f) = \Omega_p \left( \frac{f}{f_{\text{ref}}} \right)^p.
\end{equation}

\begin{figure}[h]
    \centering
    \includegraphics[width=\linewidth]{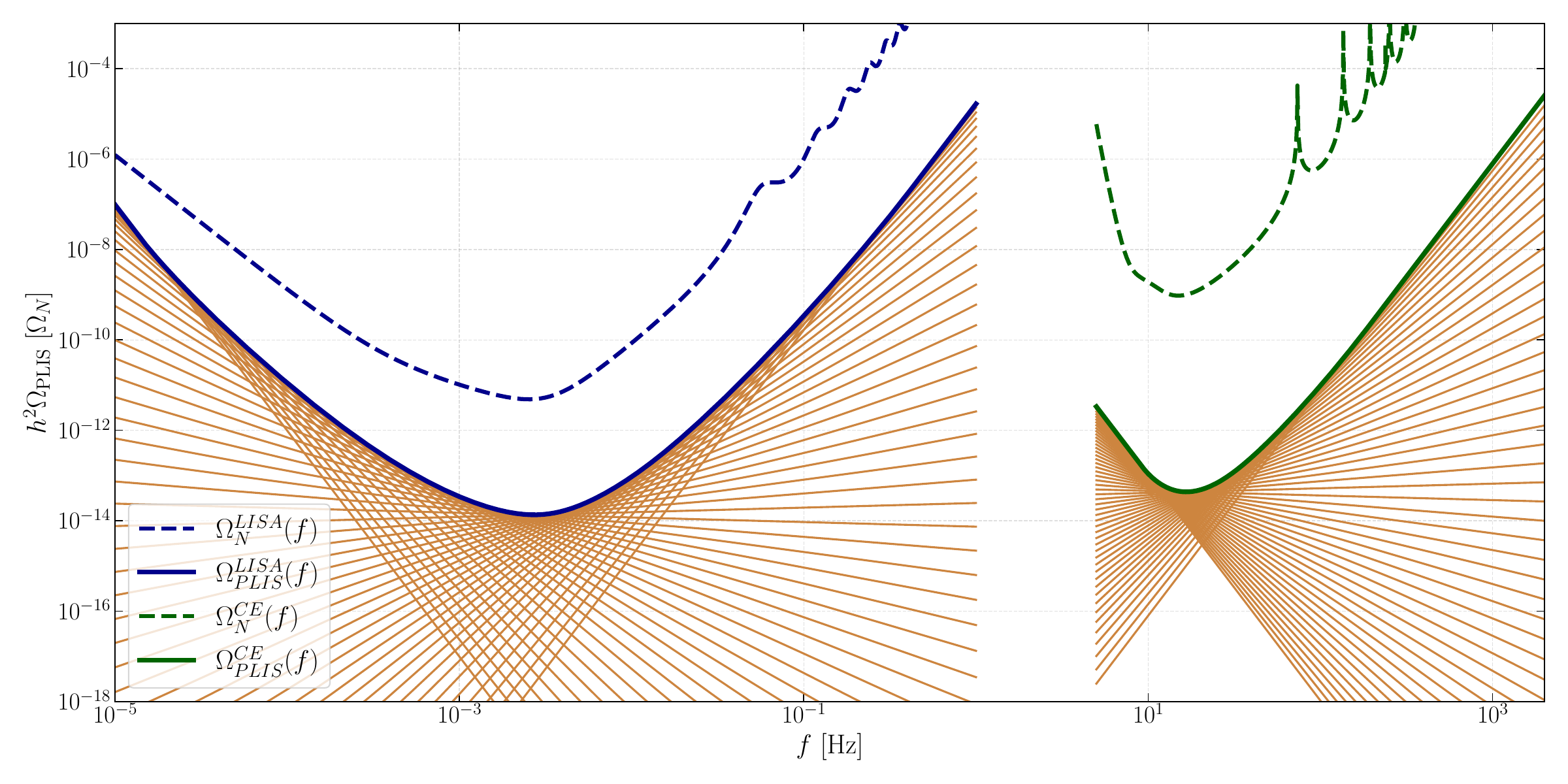}
    \caption{$\Omega_N$ and $\Omega_{\rm PLIS}$ for LISA and CE for $T=3$ yrs and $\varrho_{\text{thr}}=2$.}
    \label{fig:sens}
\end{figure}

To achieve a specific (SNR), $\varrho_{\text{thr}}$, the required amplitude $\Omega_p$ for a given power index $p$ can be calculated. The expression for $\Omega_p$ is as follows \cite{moore2014gravitational,schmitz2020new}:
\begin{equation}
    \Omega_p = \Omega_{\text{PLIS}}^{(p)} (\varrho_{\text{thr}}, t_{\text{obs}}) = \varrho_{\text{thr}} \left[ n_{\text{det}} t_{\text{obs}} \int_{f_{\text{min}}}^{f_{\text{max}}} df \left( \frac{(f/f_{\text{ref}})^p}{\Omega_{\text{N}}(f)} \right)^2 \right]^{-1/2}.
\end{equation}

where $\Omega_{\rm N}(f)$ was defined in Eq. \eqref{omega_n}. By substituting the derived $\Omega_p$ values into the Eq. \ref{eq:signaleq}, we can generate a family of power-law curves. The PLIS Curve is defined as the upper envelope of these curves Fig. \ref{fig:sens}, which is found by taking the maximum value across all possible power indices $p$.
\begin{equation}
\Omega_{\text{PLIS}}(f) = \max_p \left\{ \Omega_{\text{PLIS}}^{(p)} (\varrho_{\text{thr}}, t_{\text{obs}}) \left( \frac{f}{f_{\text{ref}}} \right)^p \right\}.
\end{equation}

To align with the Bayesian analysis in Eq.~\ref{bayesian}, we adopt a threshold $\varrho_{\text{thr}}=2$ \cite{FitzAxen:2018vdt}, which is roughly equivalent to the 95\% confidence-level limit from the likelihood analysis.

\section{GW Spectrum of Cosmic String Network \label{app:CosmicStringSpectrum}}
This appendix provides details on the string part of the GW spectrum. As Ref.~\cite{Bao:2024bws} discussed, the string-wall network produces various hybrid defects, and some of them have both a wall-mode and string-mode oscillation. The string-mode oscillation, along with the remaining stable strings after the wall network collapse, jointly contribute to the string part of the GW spectrum of the network. These hybrid defects may remain in a scaling regime by breaking off loop-like objects called cosmic rings, and the string-mode oscillation of cosmic rings can produce a GW signal. This process is analogous to the GW production of cosmic string network; thus, it is expected that, up to $\sim \order{1}$ difference, the string part of the spectrum for domain walls bounded by inflated cosmic strings is parametrically similar to that of pure cosmic strings, which has been well studied. Hence, we shall quote the benchmark cosmic string GW spectra from previous studies.

\subsection{GW Spectrum of Gauge Strings \label{app:GaugeStringSpectrum}}
We start with the gauge cosmic strings, which has been extensively studied previously \cite{Vilenkin:1981iu, Vilenkin:1981zs, Hogan:1984is, Sakellariadou:1990ne, Damour:2000wa, Siemens:2006yp, Lorenz:2010sm, Olmez:2010bi, Sousa:2013aaa, LIGOScientific:2013tfe, Stott:2016loe, Blanco-Pillado:2017oxo, Ringeval:2017eww, Cui:2017ufi, LIGOScientific:2017ikf, Cui:2018rwi, Cui:2019kkd, Blasi:2020mfx, Ellis:2020ena, Sousa:2020sxs, Figueroa:2020lvo, Co:2021lkc, Gorghetto:2021fsn, Buchmuller:2021mbb, Chang:2021afa, Gouttenoire:2021wzu, Gouttenoire:2021jhk, Boileau:2021gbr, Ferrer:2023uwz, Auclair:2023brk, Baeza-Ballesteros:2023say, Fedderke:2025sic}. The dominant GW contribution from cosmic strings is from closed string loops produced by the reconnection (or intercommutation) of long strings of horizon size. The presence of $\sim \order {1}$ long cosmic strings at all times, known as the scaling solution, warrants a one-scale model that describes how the evolution of the characteristic length of cosmic strings follows an attractor solution. Combining the one-scale model with how frequently cosmic string loops fragment from reconnecting long strings using numerical simulations, one may determine the abundance of string loops and their gravitational wave contributions. For our study, we used a particular analytical form of the gauge string spectrum provided in Ref.~\cite{Sousa:2020sxs}. Here, the sum of the two main parts gives rise to the entire GW spectrum from the gauge string,
\begin{equation}
    \Omega^{(1)}_\text{GW, str} = \Omega_\text{str, r} + \Omega_\text{str, m}.
\end{equation}
The first term $\Omega_\text{str, r}$ comes from the string loops that are produced and decay during the radiation-dominated epoch, which takes the form
\begin{equation}
    \Omega_\text{str, r}(f) = \frac{128}{9} \pi A_r \Omega_r \frac{G \mu}{\epsilon_r} \qty[ \qty( \frac{f (1 + \epsilon_r)}{B_r \Omega_m/\Omega_r + f} )^{3/2}  - \qty( \frac{f (1 + \epsilon_r)}{B_i + f} )^{3/2} ]
\end{equation}
in which 
\begin{equation}
    \begin{gathered}
        B_r \defeq \frac{4H_0\Omega_r^{1/2}}{\Gamma G \mu}, \quad 
        \epsilon_r \defeq \frac{\alpha_r \xi_r}{\Gamma G \mu}, \quad 
        B_i \defeq \frac{2}{\Gamma} \sqrt{\frac{2 H_0 \Omega_r^{1/2}}{t_i (\epsilon_r + 1)}}, \quad 
        t_i \defeq \frac{(G \mu)^2}{H_\text{re}}, \\
        \Gamma = 50, \quad 
        \alpha_r = 0.33, \quad 
        \xi_r = 0.271, \quad 
        A_r = 0.54 \mathcal{F}, \quad
        \mathcal{F} = 0.1.
    \end{gathered}
\end{equation}
Some of these parameters are extracted from numerical simulations and taken from Refs.~\cite{Sousa:2020sxs, Blanco-Pillado:2024aca}. 

Note that $t_i$, denoting the initial Hubble time at which the string is produced, in our scenario is mainly controlled by $H_\text{re}$ when the boundary strings re-enter the comoving horizon and start to oscillate. The second term $\Omega_\text{str, m}$ comes from string loops that are produced in the radiation-dominated epoch and decay in the matter-dominated epoch, which takes the form
\begin{equation}
    \begin{aligned}
        \Omega_\text{str, m}(f) =& 32 \sqrt{3} \pi (\Omega_m \Omega_r)^{3/4} H_0 \frac{A_r}{\Gamma} \frac{(\epsilon_r + 1)^{3/2}}{f^{1/2} \epsilon_r} \\
        & \times \left\lbrace \frac{(\Omega_m / \Omega_r)^{1/4}}{(B_m (\Omega_m / \Omega_r)^{1/2} + f)^{1/2}} \left[ 2 + \frac{f}{B_m (\Omega_m / \Omega_r)^{1/2} + f} \right] \right. \\
        & \left. - \frac{1}{(B_m + f)^{1/2}} \left[ 2 + \frac{f}{B_m + f} \right] \right\rbrace,
    \end{aligned}
\end{equation}
in which 
\begin{equation}
    B_m \defeq \frac{3H_0 \Omega_m^{1/2}}{\Gamma G \mu}.
\end{equation}
Following Ref.~\cite{Sousa:2020sxs}, we focus on the contribution from the fundamental mode of the cosmic string loops. Additional features on the string loops, such as kinks and cusps, may modify the UV tail of the string spectrum and higher harmonic modes, which is worth further investigation.
\begin{figure}[h]
    \centering
    \includegraphics[width=0.49\linewidth]{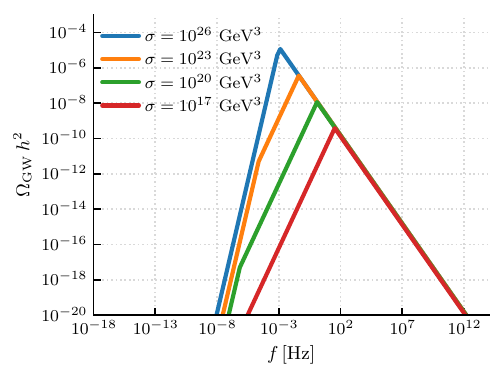}\hfill
    \includegraphics[width=0.49\linewidth]{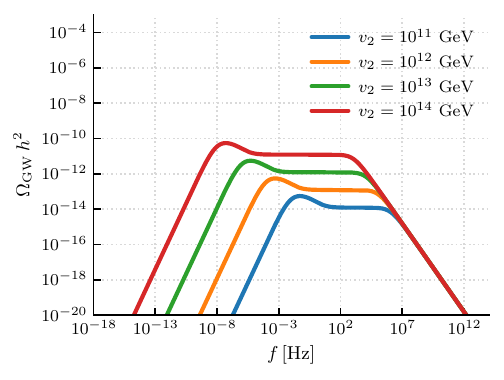}
    \caption{Exemplary spectrum for Gauge Walls and Strings with $H_{\rm re} = 10^{-10}\rm \ GeV$.}
    \label{fig:ex1}
\end{figure}

\subsection{GW Spectrum of Global Strings \label{app:GlobalStringSpectrum}}
Much similar to the gauge string, global cosmic string networks may also follow a scaling solution due to string-loop fragmentation. However, with the absence of a gauge field, the NGB becomes physical. Unlike the gauge string case, in which both the string-forming scalar field and the gauge field configurations together determine the string profile and the emission of NGBs is not possible, the spatial profile of a global string involves just the string-forming field, and its oscillation can emit NGBs. In fact, this has been considered as a possible mechanism to produce axion dark matter if the global symmetry is the Peccei-Quinn symmetry. This emission is more efficient in damping the network's energy than the GW emission. In other words, the conversion into GW is less efficient than that into NGBs, and the GW spectrum is suppressed in this case. Also, due to the absence of the gauge field, the energy density of a string becomes logarithmically divergent, which is usually cutoff by the Hubble horizon size at the IR.%
\footnote{The UV cutoff is set by the string core size, which is the Compton wavelength of the string-forming scalar $m_\phi^{-1} \sim v_2^{-1}$ assuming $\order{1}$ quartic coupling.}
The log dependence of the Hubble horizon implies that the string tension of a global string is mildly dependent on the background cosmology. How this log dependence affects the string network evolution and whether the scaling regime is violated are under active investigation, and this effect may mildly affect our conclusion on the string tension $v_2$ here. 

In this study, we used a slightly modified version of the approximate spectrum provided in Eq (3.7) of Ref.~\cite{Chang:2021afa}
\begin{equation}
    \Omega^{(2)}_\text{GW, str}(f) = 8.8\times10^{-18} \qty(\frac{v_2}{\SI{E15}{\GeV}})^4 \times 
    \begin{dcases}
        0, & f < f_0, \\
        \mathsf{L}(f_\text{eq}) \qty(\frac{f}{f_\text{eq}})^{-1/3}, & f_0 \leq f \leq f_\text{eq},\\
        \mathsf{L}(f), & f_\text{eq} \leq f < f_\eta,\\
        \mathsf{L}(f_\eta) \qty(\frac{f}{f_\eta})^{-1}, & f \geq f_\eta. \\
    \end{dcases}
\end{equation}
in which 
\begin{equation}
    \begin{gathered}
        \mathsf{L}(f) \defeq \ln[3]( \qty(\frac{2}{\alpha f})^2 \frac{v_2}{t_\text{eq}} \frac{a_\text{eq}^2}{\sqrt{\xi}} ), \quad f_0 = \SI{3.6E-16}{\Hz}, \quad f_\text{eq} = \SI{1.8E-7}{\Hz}, \\ 
        f_\eta = \frac{2}{\alpha} a_\text{eq} \sqrt{\frac{H_\text{re}}{t_\text{eq}}}, \quad
        \alpha = 0.1, \quad \xi = 4, \quad a_\text{eq} = \frac{\Omega_r}{\Omega_m}, \quad t_\text{eq} = \frac{\Omega_r^{3/2}}{\sqrt{2} \Omega_m^{2} H_0} 
    \end{gathered}
\end{equation}
Our implementation contains two minor adjustments from that in Ref.~\cite{Chang:2021afa}. First, for consistency with the gauge string spectrum quoted in \cref{app:GaugeStringSpectrum}, we dropped the dependence on the change of the number of effective degrees of freedom $\Delta_R(f)$. Correspondingly, one must match the IR and UV parts of the spectrum again so that the GW spectrum remains continuous. Second, we adjusted the UV part of the spectrum to $\sim f^{-1}$ instead of $\sim f^{-1/3}$ as this power law is attributed to higher-harmonic effects such as kinks and cusps on the string loop, which consistently dropped in \cref{app:GaugeStringSpectrum}. Dropping these factors is advantageous for the sake of consistency and numerical efficiency for now, but if a more accurate string GW spectrum can be efficiently generated for MCMC, these factors can and should be incorporated in the future.
\begin{figure}[h]
    \centering
    \includegraphics[width=0.49\linewidth]{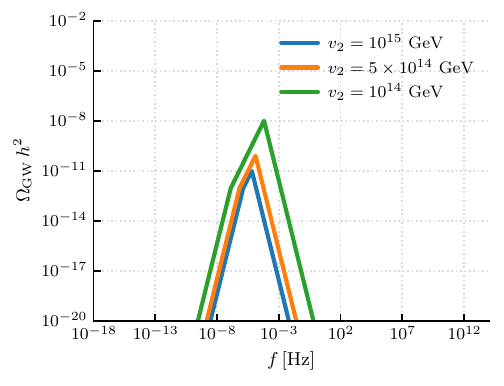}\hfill
    \includegraphics[width=0.49\linewidth]{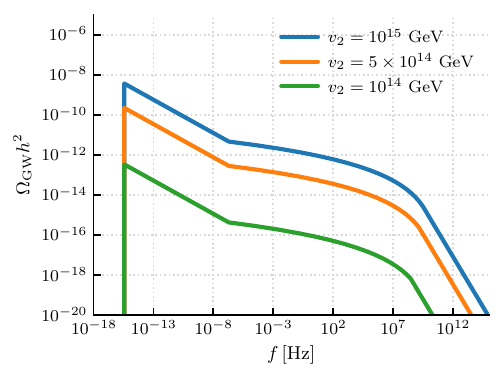}
    \caption{Exemplary spectrum for Global Walls and Strings. \textit{Left:} $H_{\rm re}=10^{-15} \rm \ GeV$, $\sigma=10^{18} \rm \ GeV^3$. \textit{Right:} $H_{\rm re}=10^{18} \rm \ GeV$.}
    \label{fig:ex2}
\end{figure}

\section{Astrophysical Foreground: Supermassive Black Hole Binaries in nHz \label{app:S_SMBHB}}
The leading contribution to GWB in the nHz frequency band is believed to be the contribution from SMBHBs. The formation of these binaries follows an evolutionary sequence that begins after a galaxy merger, proceeds through orbital decay via dynamical friction and stellar scattering, and culminates in a final inspiral phase driven by the emission of GW's \cite{burke2019astrophysics,Sesana:2013wja}. 

For analysis, the energy density spectrum of this background, $\Omega^{\text{SMBHB}}_{\text{GW}}(f)$, is modeled as a power law. This function depends on a characteristic amplitude, $A_{\text{SMBHB}}$, and a spectral index, $\gamma_{\text{SMBHB}}$. The specific functional form is:
\begin{equation}
h^2\Omega_{\text{SMBHB}}(f) = \frac{2\pi^2 A^2_{\text{SMBHB}}}{3H_0^2 }f_{\rm yr}^2 \left(\frac{f}{f_{\rm yr}}\right)^{5-\gamma_{\text{SMBHB}}} 
\end{equation}
with $f_{\rm yr} = 1/\rm yr$. For this work, we adopt a physically-informed prior derived from SMBHB population models \cite{NANOGrav:2023hfp,afzal2023nanograv}. The joint prior on $(\log_{10} A_{\text{SMBHB}}, \gamma_{\text{SMBHB}})$ is a multivariate normal distribution defined by the following mean vector and covariance matrix:
\begin{equation}
\mu_{\text{SMBHB}} = 
\begin{pmatrix} 
-15.61 \\ 
4.71 
\end{pmatrix}, 
\quad 
\sigma_{\text{SMBHB}} = 
\begin{pmatrix} 
0.2787 & -0.00263 \\ 
-0.00263 & 0.1241 
\end{pmatrix}
\end{equation}

This informed prior is used instead of broad, uninformative priors (e.g., uniform distributions) to incorporate existing knowledge from astrophysical models based on the \texttt{GWOnly-Ext} library \cite{afzal2023nanograv}. The spectral index, $\gamma_{\text{SMBHB}}$, can be compared to the theoretical benchmark of $\gamma_{\text{SMBHB}} = 13/3$, which is predicted for an idealized population of binaries evolving solely through gravitational-wave emission \cite{Phinney:2001di}. Deviations from this value can constrain the role of environmental effects in binary evolution.

\section{Full Triangle Plots and Other Best-fit Figures\label{app:full_posteriors}}
In this appendix, we show the full triangle plots with the SMBHB parameters and various GW signal plots, including the best-fit band to the NANOGrav signal of various models. 

\begin{figure}[h]
    \centering
    \includegraphics[width=\textwidth,
                     height=0.80\textheight,
                     keepaspectratio]{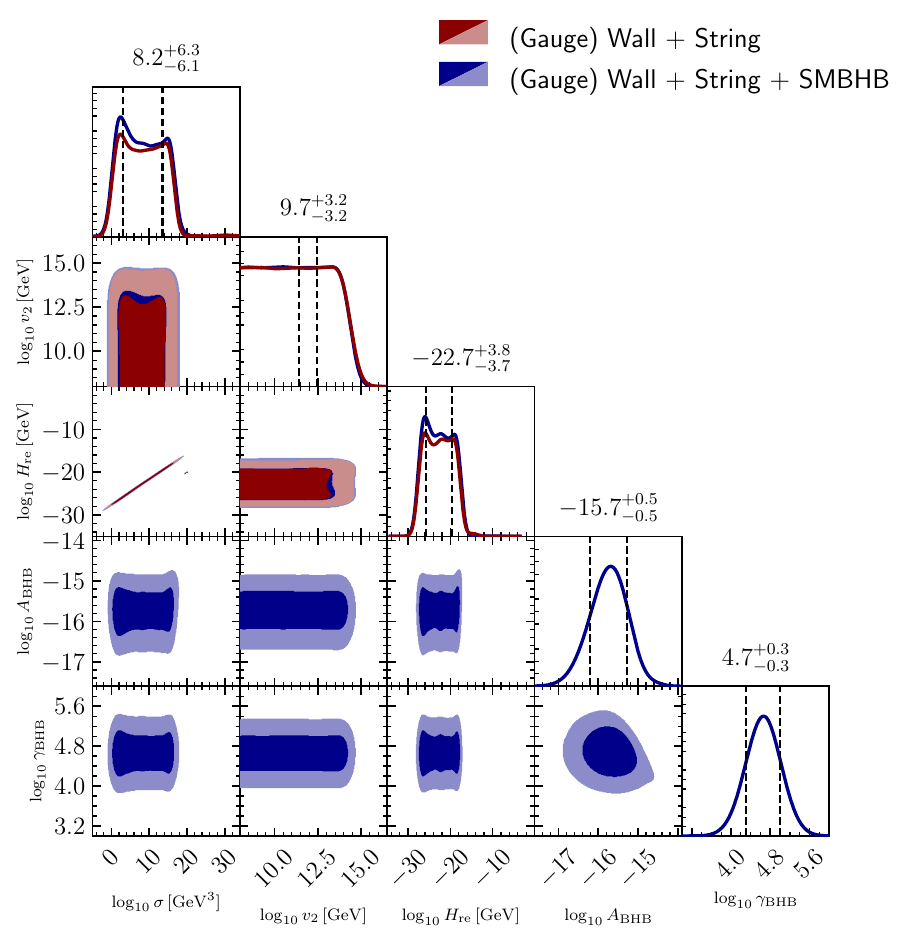}
    \caption{\textbf{New Physics Scenario 1.} Full triangle (corner) plot for the “(Gauge) Wall + String” model, with (blue) and without (red) an SMBHB foreground. Shaded contours mark the 68\% and 95\% credible regions; diagonal panels show the 1D marginals for $(\log_{10}H_{\rm re},\,\log_{10}\sigma,\,\log_{10}v_2)$ and, when present, the SMBHB parameters $(\log_{10}A_{\rm BHB},\,\gamma_{\rm BHB})$.}
    \label{fig:triangle_two_chains_gauge}
\end{figure}

\begin{figure}[h]
    \centering
    \ContinuedFloat
    \includegraphics[width=\textwidth,
                     height=0.80\textheight,
                     keepaspectratio]{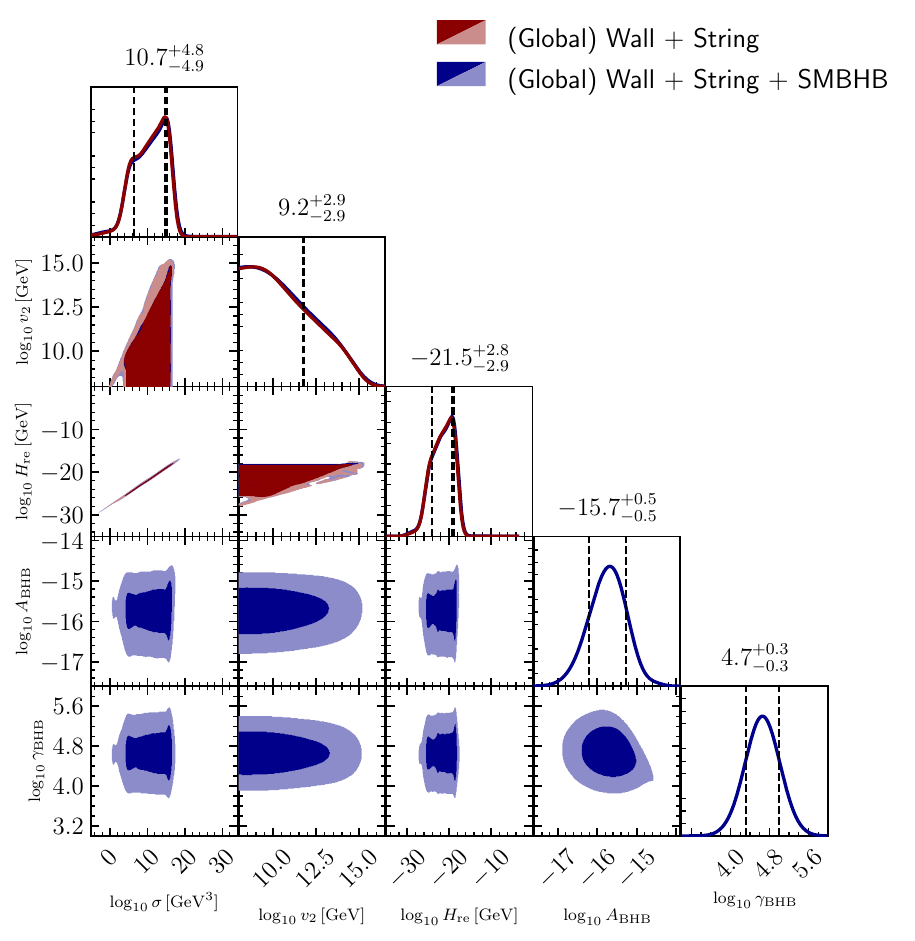}
    \caption{\textbf{New Physics Scenario 2.} Same as (a) but for the “(Global) Wall + String” model, comparing fits with (blue) and without (red) an SMBHB foreground.}
    \label{fig:triangle_two_chains_global}
    \label{fig:two_triangles}
    \caption*{}
\end{figure}

\begin{figure}
    \centering
    \includegraphics[width=\linewidth]{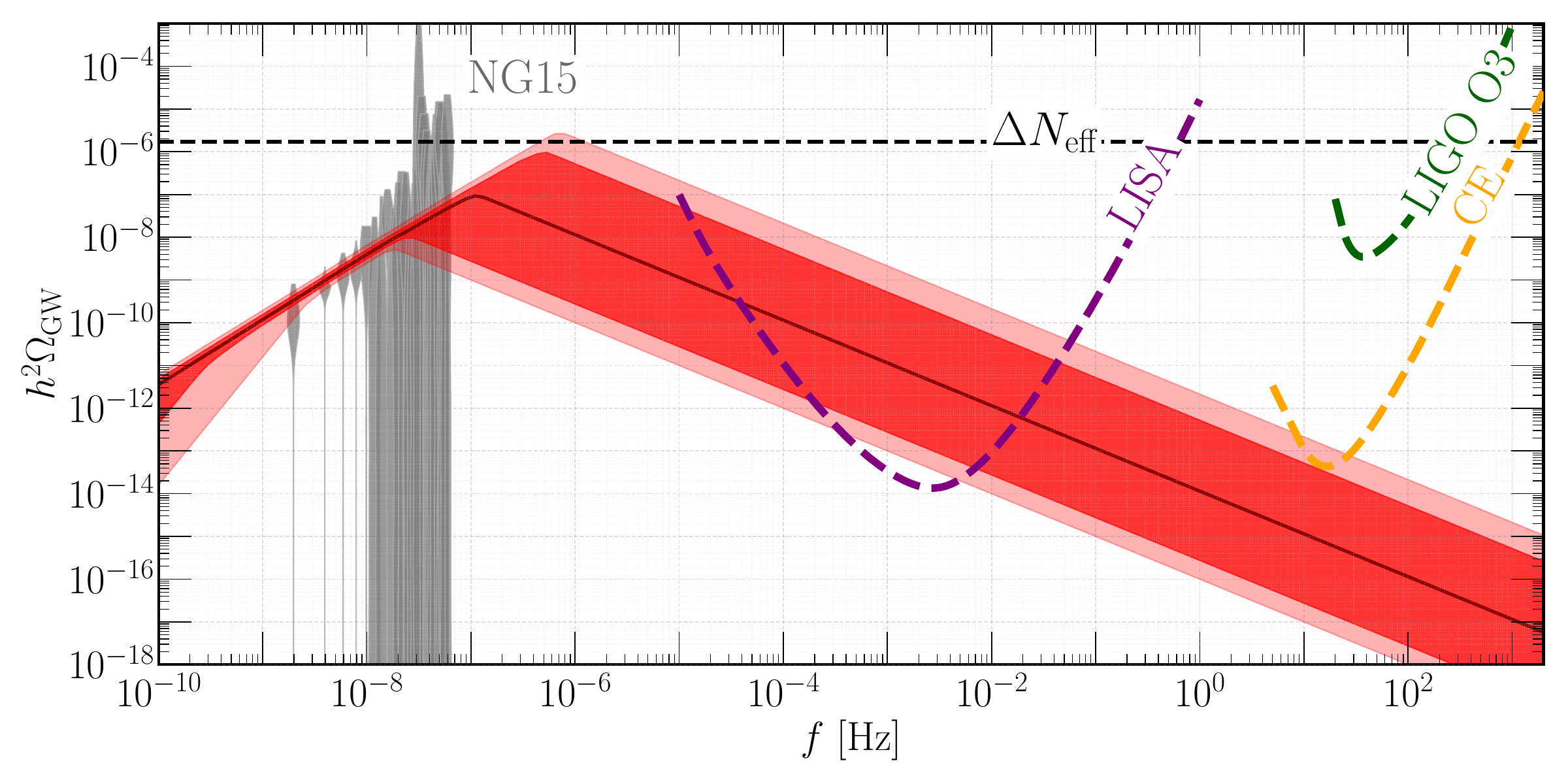}
    \caption{Posteriors of the Domain Wall part of the spectrum for the "(Gauge) Wall + String" Model. Shaded contours mark the 68\% and 95\% credible regions and the solid part is the median.}
    \label{fig:post_1}
\end{figure}

\begin{figure}
    \centering
    \includegraphics[width=\linewidth]{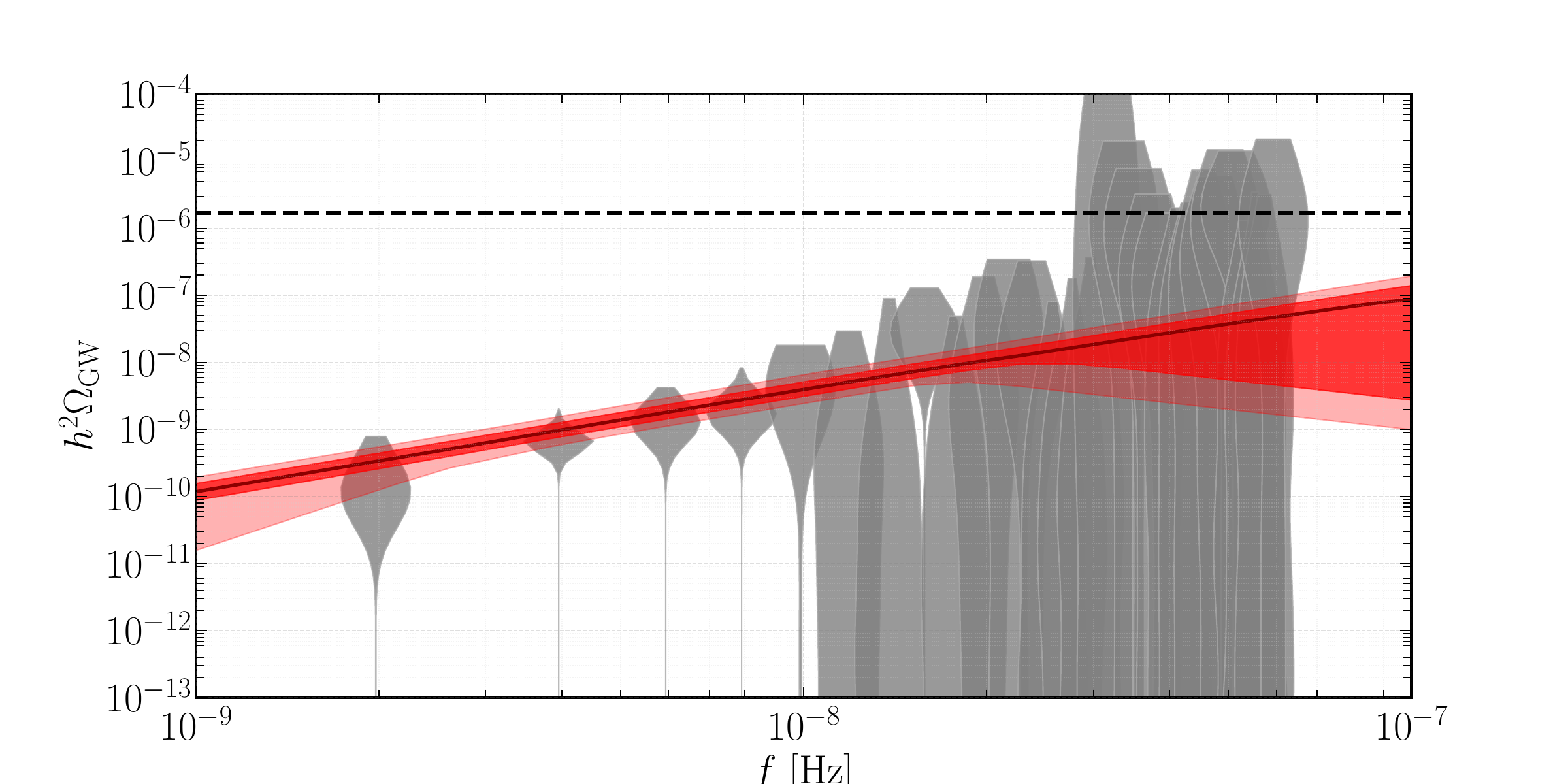}
    \caption{Close up of the Fig. \ref{fig:post_1}.}
    \label{fig:post_1_close}
\end{figure}

\begin{figure}
    \centering
    \includegraphics[width=\linewidth]{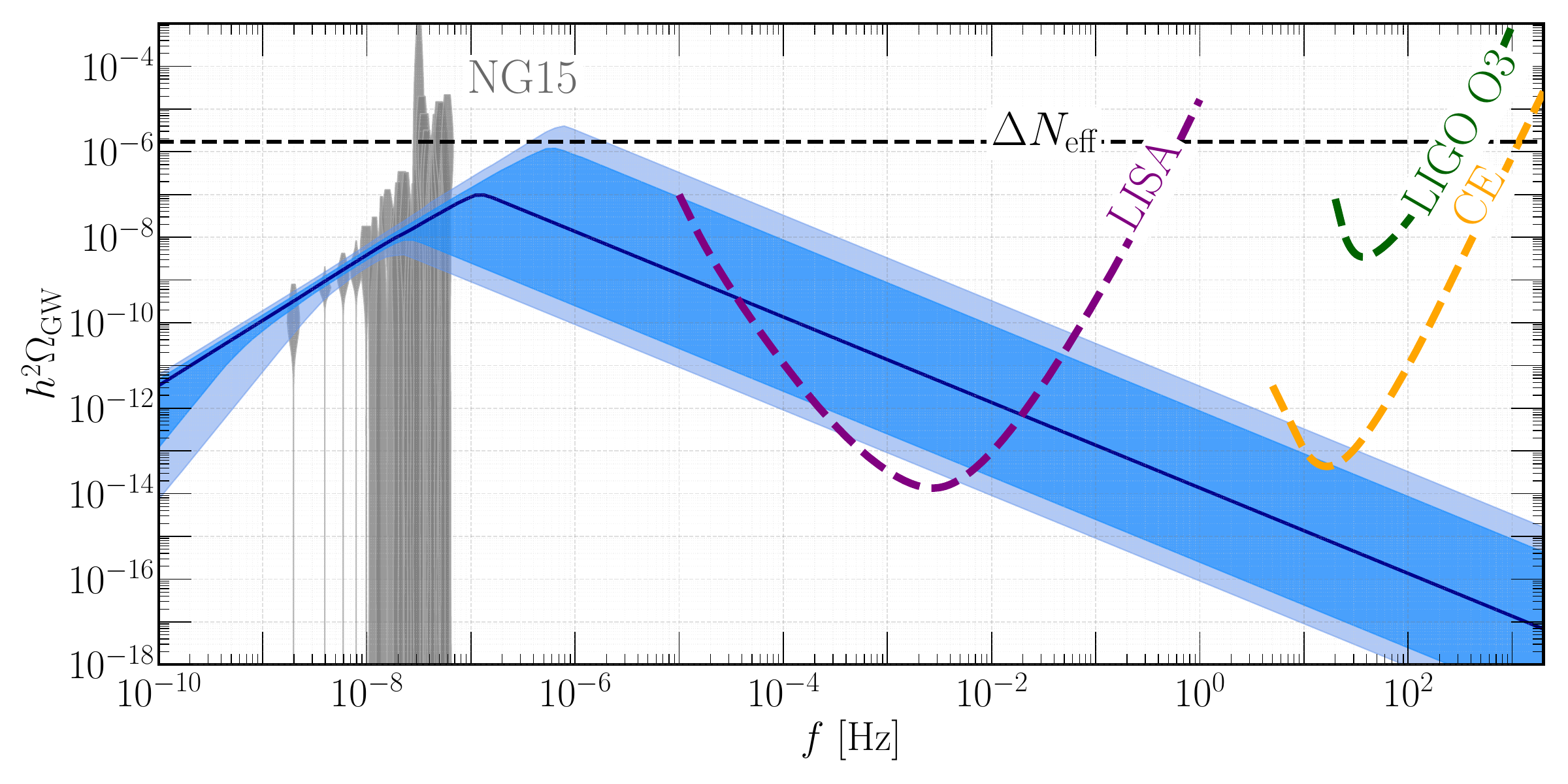}
    \caption{Posteriors of the Domain Wall part of the spectrum for the "(Gauge) Wall + String + SMBHB" Model.}
    \label{fig:post_2}
\end{figure}

\begin{figure}
    \centering
    \includegraphics[width=\linewidth]{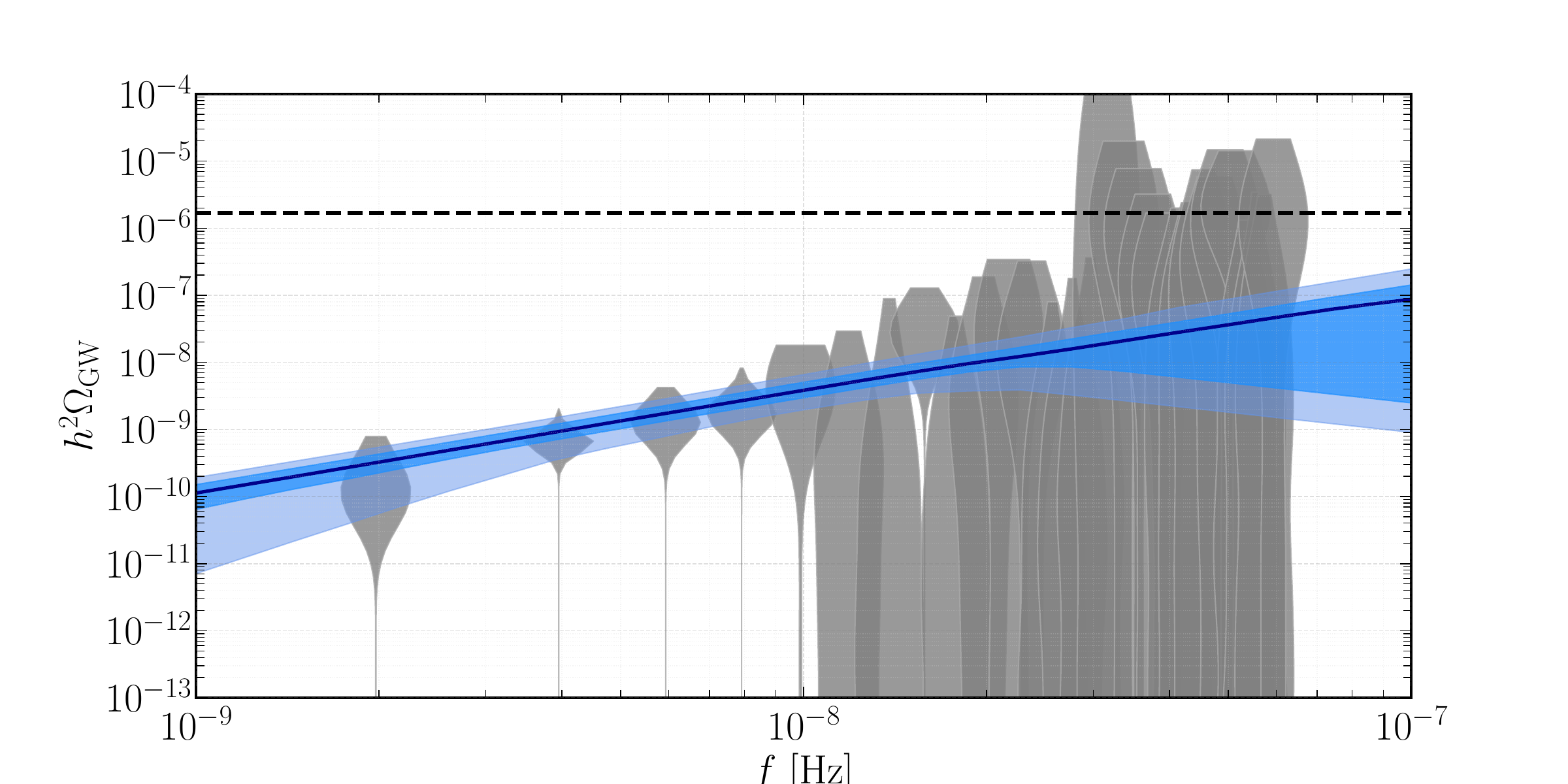}
    \caption{Close up of the Fig. \ref{fig:post_2}.}
    \label{fig:post_2_close}
\end{figure}

\begin{figure}
    \centering
    \includegraphics[width=\linewidth]{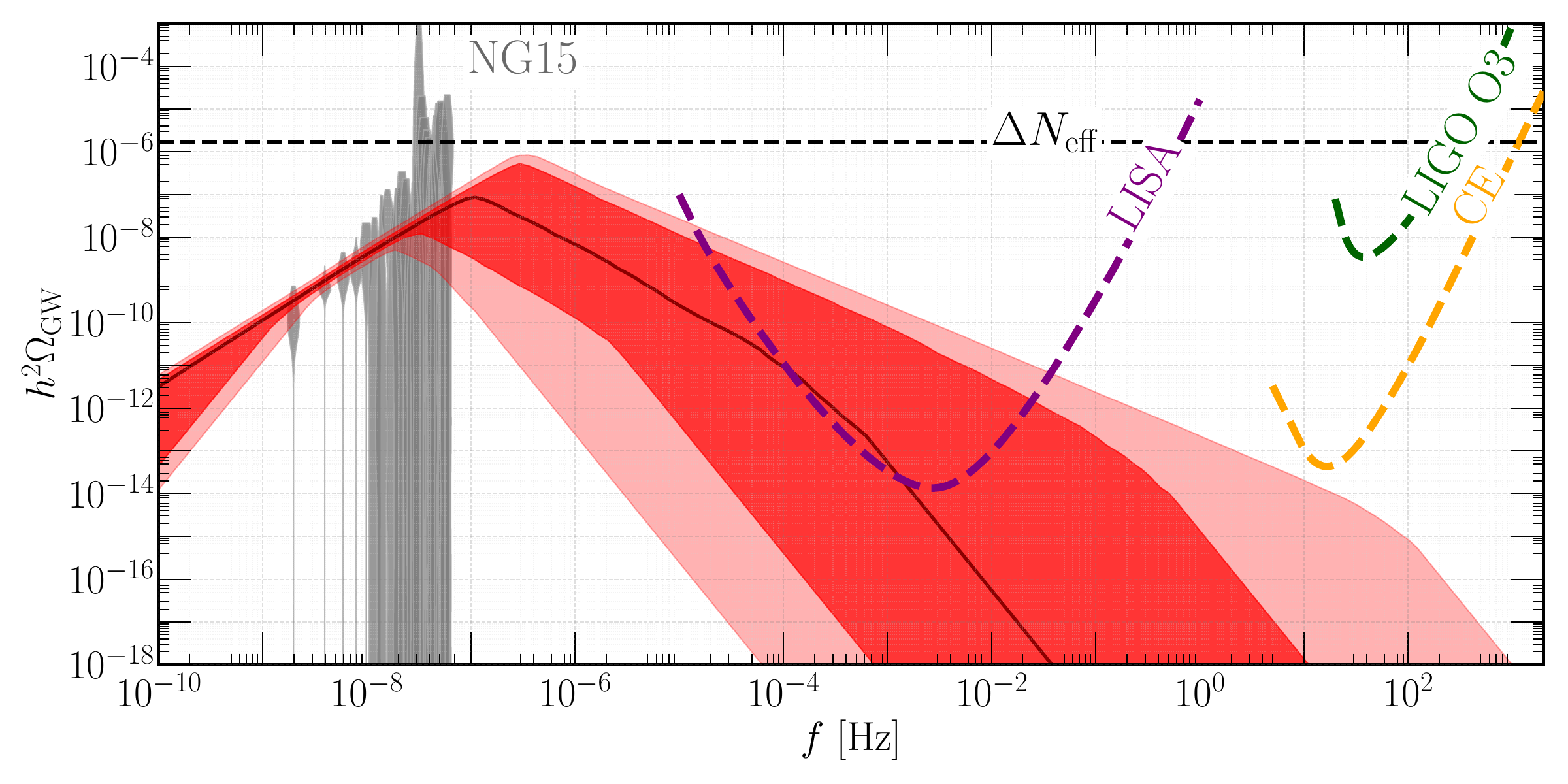}
    \caption{Posteriors of the Domain Wall part of the spectrum for the "(Global) Wall + String" Model. Shaded contours mark the 68\% and 95\% credible regions and the solid part is the median.}
    \label{fig:post_1_global}
\end{figure}

\begin{figure}
    \centering
    \includegraphics[width=\linewidth]{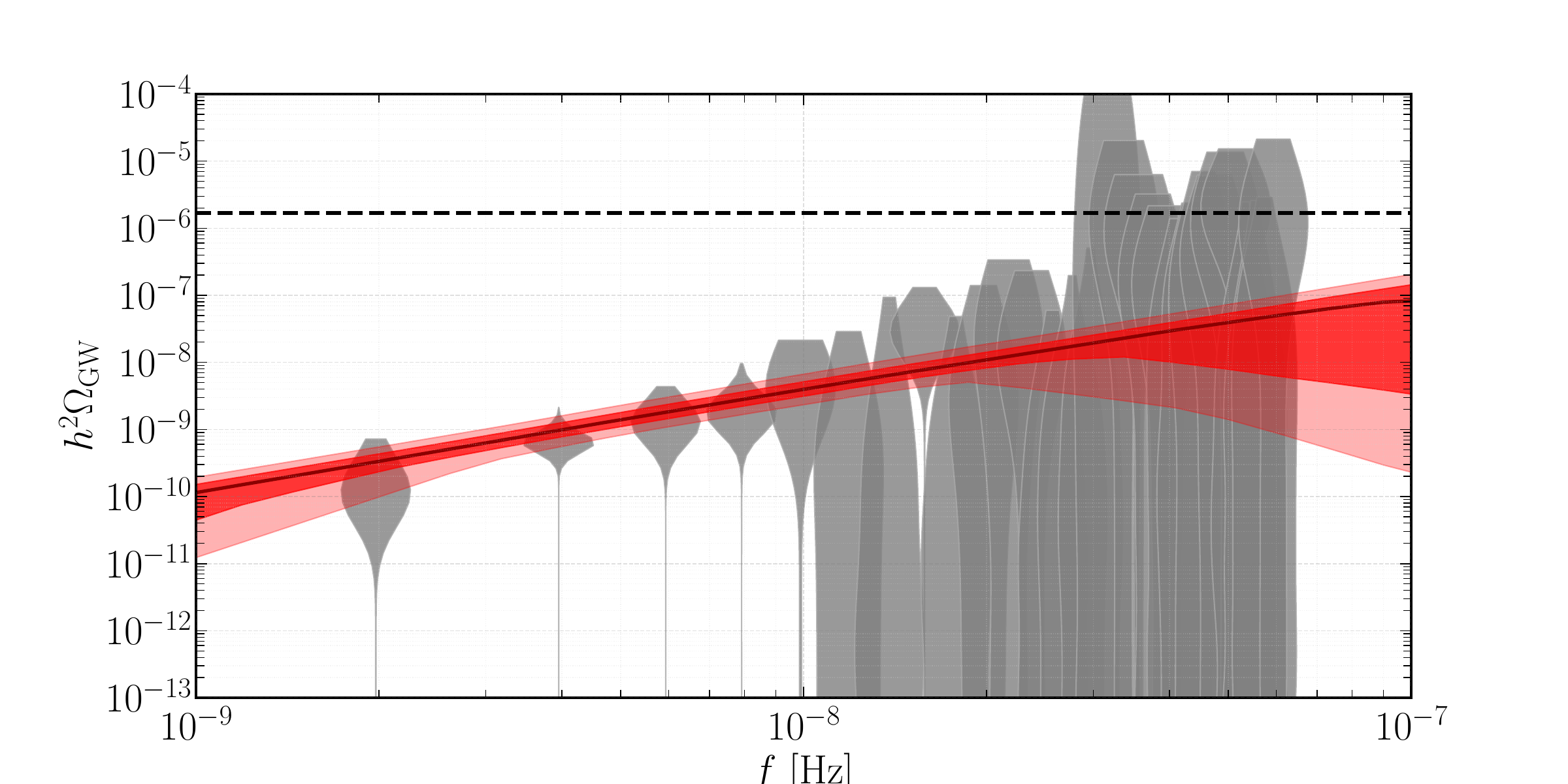}
    \caption{Close up of the Fig. \ref{fig:post_1_global}.}
    \label{fig:post_1_close_global}
\end{figure}

\begin{figure}
    \centering
    \includegraphics[width=\linewidth]{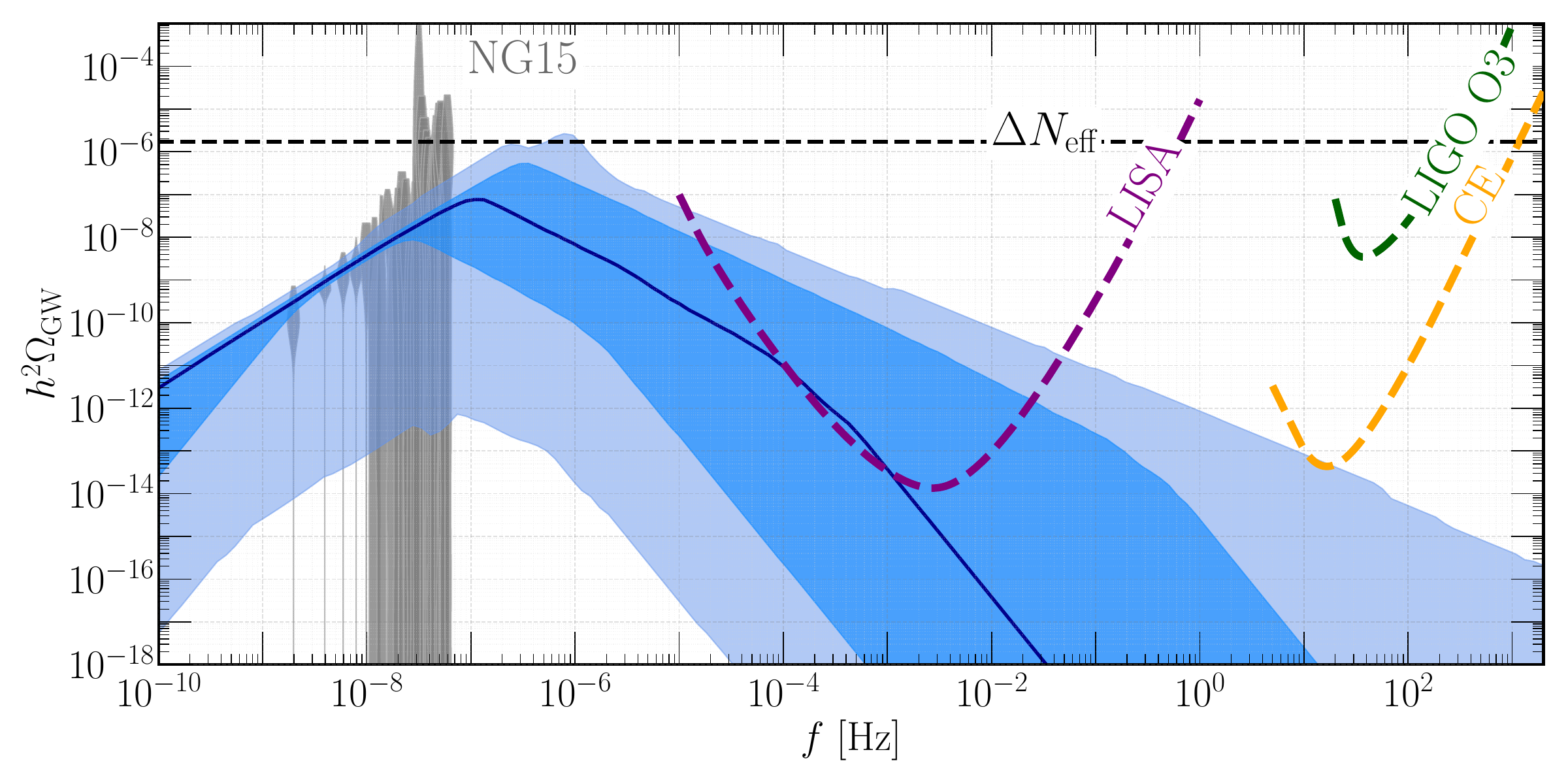}
    \caption{Posteriors of the Domain Wall part of the spectrum for the "(Global) Wall + String + SMBHB" Model.}
    \label{fig:post_2_global}
\end{figure}

\begin{figure}
    \centering
    \includegraphics[width=\linewidth]{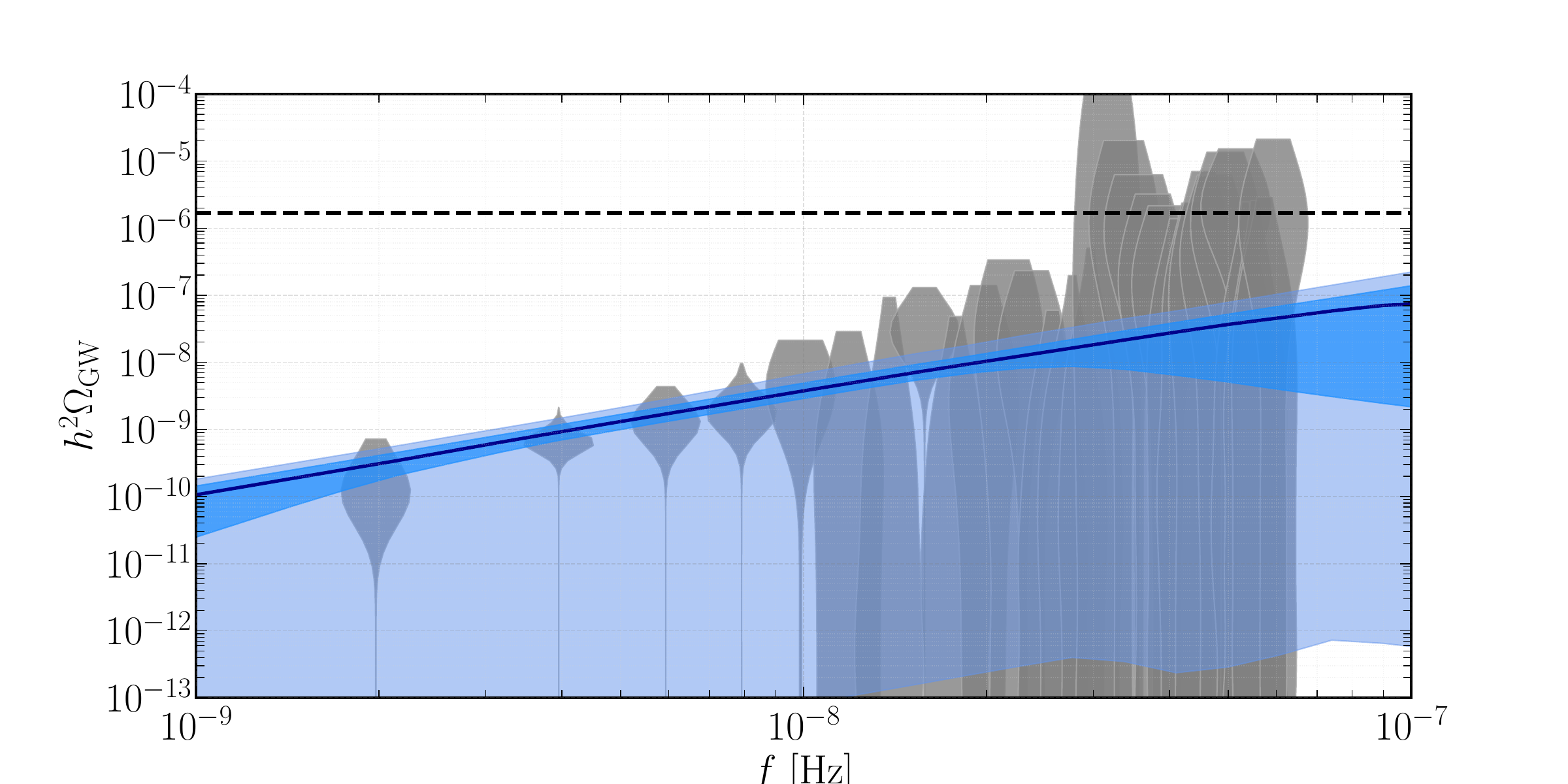}
    \caption{Close up of the Fig. \ref{fig:post_2_global}.}
    \label{fig:post_2_close_global}
\end{figure}

\end{document}